# Large time off-equilibrium dynamics of a particle in a random potential

Leticia F. Cugliandolo

*Service de Physique de L'Etat Condensé, Saclay, CEA*

*Orme des Merisiers, 91191, Gif–sur–Yvette Cedex, France.*

e-mail: leticia@amoco.saclay.cea.fr

Pierre Le Doussal

*CNRS-Laboratoire de Physique Theorique de l'Ecole Normale Superieure,*

*24 rue Lhomond, 75231 Cedex 05, Paris, France.*

e-mail: ledou@physique.ens.fr

(May 24, 1995)

# Abstract

We study the off-equilibrium dynamics of a particle in a general $N$-dimensional random potential when $N \to \infty$. We demonstrate the existence of two asymptotic time regimes: *i.* stationary dynamics, *ii.* slow aging dynamics with violation of equilibrium theorems. We derive the equations obeyed by the slowly varying part of the two-times correlation and response functions and obtain an analytical solution of these equations. For short-range correlated potentials we find that: *i.* the scaling function is non analytic at similar times and this behaviour crosses over to ultrametricity when the correlations become long range, *ii.* aging dynamics persists in the limit of zero confining mass with universal features for widely separated times. We compare with the numerical solution to the dynamical equations and generalize the dynamical equations to finite $N$ by extending the variational method to the dynamics.

# I. INTRODUCTION

A major theoretical challenge, as well as an important issue for numerous experimental systems, is to understand the non-equilibrium dynamics of elastic manifolds, with or without internal periodic structure, in quenched random media[1]. The corresponding models, such as the model of an interface subjected to quenched impurities or the Sine Gordon model with phase randomness, have been studied for some time. Some progress has been made in the description of the statics[2-6] but little is still known about the dynamics, especially about the non-equilibrium features of the relaxations. It is by now well-established that these systems exhibit glassy behaviour which presents similarities with the glassiness of spin-glasses, such as slow relaxations, history dependence and strong sensitivity to changes in the external parameters[7,8].

Recently progress has been made in solving the non-equilibrium dynamics of mean-field models of spin-glasses[9-13]. A new method to study analytically the large-times off-equilibrium dynamics has been introduced[10,11]. This purely dynamical method has on the one hand some formal connections with the replica symmetry breaking schemes introduced to study the static properties, *i.e.* the Gibbs measure, of these models. On the other hand, the method yields interesting additional information specific to the dynamics. It allows us to find a solution which exhibits two asymptotic time regimes: a stationary dynamics regime for large but similar times, and a slow aging regime for widely separated times. Most importantly it then allows to establish contact with the essentially out of equilibrium experimental observations, namely the slow relaxations and the aging effects[7]. It is then natural to extend this method to study the dynamics of a broader class of systems with slow relaxations including the random manifold problem.

For problems such as the manifold in a random medium, the mean-field limit corresponds to the dimension of the embedding space $N$ going to infinity, *i.e.* it is represented by a field-theory with a large number of components. In this case one can derive a closed set of dynamical equations and the above dynamical method can be applied yielding exact results. Realistic systems, however, are embedded in a finite dimensional space, *i.e.* $N$ is finite, and one must make some approximation to obtain closed dynamical equations. A way to obtain these equations is to extend the Gaussian Variational Approximation (GVA) to the dynamics.

The model of a manifold of internal dimension $D$ embedded in a random medium of dimension $N$ is described, in terms of a $N$ component displacement field $\boldsymbol{\phi}$, $\boldsymbol{\phi} = (\phi_1, \phi_2, \ldots, \phi_N)$, by the Hamiltonian[4,5]

$$H = \int d^D x \left[ \frac{c}{2} (\nabla \phi(x))^2 + V(\phi(x), x) + \frac{\mu}{2} \phi^2 \right]. \quad (1.1)$$

$\mu$ is a mass, which effectively constraints the manifold to fluctuate in a restricted volume of the embedding space. $V$ is a Gaussian random potential with correlations

$$\overline{V(\boldsymbol{\phi}, x) V(\boldsymbol{\phi}', x')} = -N \delta^D(x - x') \, \mathcal{V}\left( \frac{(\boldsymbol{\phi} - \boldsymbol{\phi}')^2}{N} \right). \quad (1.2)$$

Our aim is to study the out of equilibrium dynamics of this model for a general random potential. The dynamics that we consider is the Langevin dynamics

$$\frac{\partial \phi(x, t)}{\partial t} = -\frac{\delta H}{\delta \phi(x, t)} + \boldsymbol{\eta}(x, t) \quad (1.3)$$

with $\langle \eta_\alpha(x, t) \eta_\beta(x', t') \rangle = 2T \, \delta_{\alpha\beta} \, \delta^D(x - x') \delta(t - t')$.

To summarize, in this paper we concentrate on the problem of a particle moving in a $N$ dimensional random potential. It corresponds to the limit of a manifold with zero internal dimension[14,15] $D = 0$. The zero dimensional case is a necessary step before studying the model in finite $D$ and a good trial ground for the method. Schematically we do the following. Firstly we study analytically the exact dynamical equations in the mean-field limit $N \to \infty$. Secondly, we derive the dynamical equations for finite $N$ within the extended GVA and we apply the same dynamical method to the approximated set of equations. We shall present the analysis of the finite dimensional models ($D > 0$) with the necessary extensions of the method in a separate publication[16].

The quantities of interest in the long-time off-equilibrium dynamics are the two time correlation and response functions. In the $D = 0$ model we define

$$C(t,t') = \overline{\langle \phi(t)\phi(t') \rangle} \qquad R(t,t') = \overline{\frac{\delta \langle \phi(t) \rangle}{\delta f(t')}}\bigg|_{f=0} \qquad (1.4)$$

where $f(t')$ is a small perturbation applied at time $t'$. For the models we consider here it is also convenient to use the mean squared displacement correlation function $B(t,t') = C(t,t) + C(t',t') - 2C(t,t') = \overline{\langle (\phi(t) - \phi(t'))^2 \rangle}$.

Infinite dimensional models $N \to \infty$

There are several models which can be studied for $N \to \infty$ and $D = 0$ corresponding to different choices of the random potential correlation $\mathcal{V}(z)$. A common choice is

$$\mathcal{V}(z) = \frac{(\theta + z)^{1-\gamma}}{2(1-\gamma)} \ . \qquad (1.5)$$

In what follows we shall refer to this model as the power law model. There are two physically distinct cases. If $\gamma < 1$ the correlations grow with the distance $z$ and the potential is called 'long-range'. If $\gamma > 1$ the correlations decay with the distance and the potential is called 'short-range'. The statics of these models has been studied with the replica trick[5,17]. The short range case is considered to be solved with a one step replica symmetric ansatz while the long-range case needs a full replica symmetry breaking scheme.

One can also define spherically constrained models letting the mass $\mu$ become a time-dependent function $\mu(t)$ related to the Lagrange multiplier enforcing the $N$-spherical constraint. Model (1.1) then represents a manifold of internal dimension $D$ embedded on the surface of a $N$-dimensional sphere and it can be related to spherically constrained spin-glass models.

In particular, the $p$-spin spherical spin-glass model defined by the Hamiltonian[18]

$$V(\phi) = \sum_{i_1 < \ldots < i_p}^{N} J_{i_1,\ldots,i_p} \phi_{i_1} \ldots \phi_{i_p} \qquad (1.6)$$

with the additional spherical constraint $\sum_{i=1}^{N} \phi_i^2 = N$ and $J_{i_1,\ldots,i_p}$ taken from a Gaussian distribution is contained in this family of models. It is recovered for $D = 0$ and the correlation of the random potential $\mathcal{V}$ given by:

$$\mathcal{V}(x) = -\frac{1}{2}\left(1 - \frac{x}{2}\right)^p . \tag{1.7}$$

The replica analysis yields an exact replica symmetric solution[19] when $p = 2$. However, this is a marginal case and an infinitesimal departure from $p = 2$ makes the exact solution be one step replica symmetry breaking[18]. The out of equilibrium dynamics has been studied in Ref.[33,13] for $p = 2$ and in Ref.[10] when $p \geq 3$.

We study these models as follows[10–13]. The thermodynamic limit $N \to \infty$ is taken initially and then, only afterwards, the limit of large times is eventually taken. The dynamics starts at a finite initial time $t_o = 0$. The initial condition is chosen at random from a Gaussian distribution. The models we study have many ergodic components and the system *is not able* to jump from one component to another since, by definition, the barriers separating different ergodic components are divergent and hence unsurmountable. However, almost all the initial conditions do not lead the system to equilibrium at large time. A clear realisation of this is observed studying the dynamics of the $p = 2$ spherical spin-glass[13] starting from a generic initial condition for which the system evolves indefinitely without ever reaching any of the two equilibrium states.

The two assumptions that describe quantitatively this scenario are the *weak ergocity breaking* hypothesis[20,10] and the *weak long-term memory* hypothesis[10,11]. The former states that even in the limit of large time the particle is always able to escape from its previous positions and never gets stuck in a local equilibrium. The latter states that at a sufficiently large time $t$ the particle forgets any small perturbation applied during any previous *finite* time interval. For instance the 'remanent magnetisation' associated to a field applied during a finite time interval $[0, t_w]$ decays to zero for a sufficiently large $t$. For mean-field spin-glass models the relaxations slow down as time passes though the systems do not reach any kind of equilibrium. There is no equilibration time $t_{eq}$ such that for all the subsequent times the 'equilibrium-dynamics' theorems, *i.e.* time-translation invariance (TTI) for all the correlation functions and the fluctuation-dissipation theorems (FDT) hold. The systems do not visit the equilibrium states at any time. Importantly enough, this scenario is what is observed during experimental times in real spin-glasses. Despite being out of equilibrium, mean-field spin-glass models as well as real spin glasses reach an asymptotic (large times) regime for which some general features can be demonstrated. Generalisations of the equilibrium theorems to the large-times out of equilibrium relaxations have been proposed in Refs.[10,11] and we shall apply them here to study model (1.1) for a general $\mathcal{V}$.

It is important to note that this scenario is completely different from the one proposed by Sompolinsky and Zippelius[21,22] to study *the equilibrium properties* of mean-field spin-glass models using a dynamical approach. In the so-called Sompolinsky-dynamics one takes a large but *finite* system (finite $N$). One then chooses the initial time to be minus infinity, in such a way that one assumes that at finite times the system has arrived at a certain equilibrium state. One namely takes the inverse order of limits as above: $\lim_{N\to\infty}\lim_{t\to\infty}$ to allow for jumping from one ergodic component to another[22]. In this approach time translational invariance is always assumed and hence aging effects are *not* captured. Moreover, the method has some problems since when $N$ is finite and times diverge with respect to $N$ quantities such as the the auto-correlation function become non self-averging[23]. The model (1.1)-(1.5) has been studied previously in this way by Kinzelbach and Horner[24,25].

In this paper we find that, in the low temperature phase, for large times $t \geq t'$ such that $(t - t')/t' << 1$ the particle undergoes stationary dynamics $B(t, t') = B_F(t - t')$ where $B_F(t - t')$ grows from 0 to the limiting value $q$. For larger time separation $(t - t')/t' = O(1)$ there is an aging regime where $B(t, t')$ further grows slowly from $q$ to $b_o$. The behaviour in the aging regime depends on the range of the correlations of the potential. For long-range correlations we find that the one time quantities converge to the values predicted by the statics. This confirms the numerical observations of Ref.[12] for model (1.1) with $\mathcal{V}$ given by (1.5) and $\gamma < 1$. We demonstrate that $B(t,t')$ satisfies ultrametricity in time, and we explicitly compute the rate of violation of the FDT theorem at large times. For short-range correlated potentials we find that one time quantities which involve the aging regime, such as the asymptotic energy density or $b_o$, differ from their equivalents in the statics. The dynamical phase diagram is different from the equilibrium one and, in particular, the transition in the dynamics survives in the limit of vanishing mass $\mu \to 0$. We find that in the aging regime $B(t,t') = \hat{B}[h(t')/h(t)]$ where generically the scaling function $\hat{B}[\lambda]$ is non analytic when $\lambda \to 1$ ($t'/t \to 1$) with a non trival exponent $\alpha$. Furthermore in the limit of vanishing mass $\mu \to 0$ we show that, for all short-range models, $\hat{B}[\lambda] \sim q + \ln(1/\lambda)$ for $\lambda \ll 1$. This remarkable form implies that the response function at large time separation becomes a function of $t'$ only. This is consistent with some numerical results that we present.

The parameter $\gamma$ in the definition of $\mathcal{V}$, Eq. (1.5), tunes from long-range correlated potentials ($\gamma \leq 1$) to short-range correlated potentials ($\gamma > 1$). Interestingly enough, this allows us to explicitly show how the ultrametric dynamical solution is approached when $\gamma \to 1$ from above.

The limit of zero mass raises some fundamental questions. For $\mu > 0$, $C(t,t)$ reaches a finite limit when $t \to \infty$. By contrast, at strictly $\mu = 0$, $C(t,t)$ grows unboundedly with time. Although the corresponding unbounded diffusion process deserves further study[16], we show that the method still applies and the limit $\mu \to 0$ can safely be taken for quantities such as $B(t,t')$ and $R(t,t')$.

A first analysis of the out of equilibrium relaxation of model (1.1) at $D = 0$ with long-range correlations given by Eq.(1.5) has been carried out by Franz and Mézard[12]. They have solved numerically the mean-field dynamical equations and have proposed a scenario of non-overlapping time domains to account for their numerical results.

Finite dimensional models $N < \infty$

The model (1.1) with finite $N$ (and $D = 1, 2, 3$) has several physically interesting realizations, such as the problem of a manifold pinned by impurities for which the correlator is of the form (1.5). It arises in the study of interfaces in a random field as well as glassy phases of vortices in high $T_c$ superconductors. Similarly the Sine Gordon model with random phase disorder (RSGM) arises in the study of several problems including quenched disorder. For example it describes the glass transition of a surface of a crystal deposited on a disordered two-dimensional substrate. It is also related to the vortex-free XY model in a random field. It is defined in term of a $N = 1$ phase field $\psi$ by:

$$H = \int d^D x \left[ \frac{c}{2} (\nabla \psi(x))^2 - \zeta_1(x) \cos(\psi(x)) - \zeta_2(x) \sin(\psi(x)) + \frac{\mu}{2} \psi^2 \right] . \tag{1.8}$$

where $\overline{\zeta_i(x)\zeta_j(x')} = \delta_{ij}\delta(x - x')$. This model corresponds to the choice of correlator:

$$\overline{V(\psi, x) V(\psi', x')} = -\delta^D(x - x') \cos(\psi - \psi') . \tag{1.9}$$

The statics of the RSGM has been studied using renormalization group techniques[2], a Gaussian approximation (variational method) complemented by the replica trick[6] and with extensive numerical simulations[3].

The statics of model (1.1) for general $D$ and $N$ has been studied using a Hartree, *i.e.* a Gaussian variational approximation[5]. The corresponding mean-field equations are identical to those for $N$ infinite, apart from a replacement of the random potential correlator $\mathcal{V}(x)$ by $\hat{\mathcal{V}}(x)$ defined as $\hat{\mathcal{V}}(\langle\phi\rangle_0^2) \equiv \langle \mathcal{V}(\phi^2)\rangle_0$ where $\langle \cdot \rangle_0$ denotes a Gaussian average over $\phi$. In this paper we extend the GVA to the dynamics by performing a Gaussian decoupling approximation in the exact dynamical equations of motion. As it is detailed in Appendix A, this shows explicitly that the same replacement of $\mathcal{V}(x)$ by $\hat{\mathcal{V}}(x)$ holds in the dynamics. For instance the RSGM dynamics is described by Eq. (1.5) with $\mathcal{V}(x) \to \hat{\mathcal{V}}(x) = -\Delta \exp(-x/2)$.

The zero dimensional version $D = 0$ has been studied as a toy model of the above problems[14,15]. Even if one does not expect strictly speaking a glass transition in $D = 0$, this model exhibits at low temperatures several features of a glass which are present in its higher dimensional versions. In the statics perturbation theory breaks down because of the large number of metastable states and the importance of rare fluctuations. In the dynamics, there is a finite ergodic time $t_{eq}$ beyond which equilibrium dynamics is established for all subsequent times. This time however can be very large since barriers grow as powers of the inverse mass, and $N$. Therefore even in this simple case, there could be an aging regime at intermediate times. If this is the case it is likely that the methods of the present paper, using the GVA, will provide a basis for describing this out of equilibrium regime.

In the limit of a zero mass, this model corresponds to the problem of diffusion in a random potential which has been extensively studied. In the particular case $\gamma = 0$ $N = 1$ model (1.1-(1.5) is the celebrated Sinai's model[26]. More generally it is known that $\phi(t) \sim \log(t)^{(2/(1-\gamma))}$ for any finite $N$. However the two time correlations and thus the aging properties have not been studied previously analytically, except in the case of an applied force in Ref Fevi (see also Ref Mapa).

To summarise, in this paper we study analytically the long time behaviour of model (1.1) with general $\mathcal{V}$ in the simplest case of zero internal dimensions. We use the formalism of Refs.[10,11] and we concentrate on the low temperature phases. We also reproduce and explain as explicitly as possible, some of the necessary calculations. The paper is organised as follows. In Section II we present the general mean-field dynamical equations for the zero internal dimension case. In Section III we describe the separation of the asymptotic dynamics in a stationary regime for long but similar times and a non-stationary regime for long and very different times. We derive the dynamical equations for both regimes. In Section IV we review the extensions of the equilibrium dynamics theorems to the asymptotic out of equilibrium relaxation. In Section V we present the - time-reparametrisation invariant - equations for the aging regime. We also discuss the strategy we follow in Sections VI and VII to find their solutions. In Section VI we use a 'one blob' ansatz to solve the short-range models and in Section VII we use an ultrametric ansatz to solve the long-range models.

Finally, in Section VIII we present our conclusions.

## II. MEAN FIELD EQUATIONS IN THE LARGE TIME LIMIT

The general dynamical equations for the two-time response function $R(t,t')$ and correlation function $C(t,t')$ for the model defined by Eq. (1.1) in the limit $N \to \infty$ were derived in Ref.[12]. For the special case of the $p$-spin spherical model they were presented in Ref.[10]. The dynamical equations *assuming* time-translation invariance (TTI) at all times for all the two-time functions were previously presented in Refs.[24],[25],[30] and [31] for model (1.1) with short and long correlation, the $p$-spin spherical spin-glass and the RSGM, respectively. In these references the equilibrium dynamics à la Sompolinsky has been studied.

In Appendix A we sketch a derivation of the general dynamical equations using the Martin-Siggia-Rose formalism complemented by a Gaussian approximation that becomes exact in the $N \to \infty$ limit. The mean-field dynamical equations in terms of $C$ and $R$ defined in Eq.(1.4) read

$$\frac{\partial R(t,t')}{\partial t} = -\mu R(t,t') + 4 \int_0^t ds \, \mathcal{V}''(B(t,s)) \, R(t,s) \, (R(t,t') - R(s,t')) \qquad (2.1)$$

$$\frac{\partial C(t,t')}{\partial t} = -\mu C(t,t') + 2 \int_0^{t'} ds \, \mathcal{V}'(B(t,s)) \, R(t',s)$$
$$+ 4 \int_0^t ds \, \mathcal{V}''(B(t,s)) \, R(t,s) \, (C(t,t') - C(s,t')) + 2T \, R(t',t) \,. \qquad (2.2)$$

We use the Ito convention $\lim_{\epsilon \to 0} R(t, t-\epsilon) = 1$, $R(t,t) = 0$.

Using the fact that the last term in Eq.(2.2) vanishes for $t > t'$ one also has

$$\frac{1}{2}\frac{dC(t,t)}{dt} = -\mu C(t,t) + T + 2 \int_0^t ds \, \mathcal{V}'(B(t,s)) \, R(t,s) +$$
$$2 \int_0^t ds \, \mathcal{V}''(B(t,s)) \, R(t,s) \, (C(t,t) - C(s,s) + B(t,s)) \,. \qquad (2.3)$$

For these problems it is convenient to use the mean squared displacement correlation function $B(t,t') = C(t,t) + C(t',t') - 2C(t,t') = \overline{\langle (\phi(t) - \phi(t'))^2 \rangle}$ that satisfies for $t > t'$

$$\frac{1}{2}\frac{\partial B(t,t')}{\partial t} = -\frac{\mu}{2} \, [C(t,t) - C(t',t') + B(t,t')] + 2 \int_0^t ds \, \mathcal{V}'(B(t,s)) \, (R(t,s) - R(t',s))$$
$$+ T + 2 \int_0^t ds \, \mathcal{V}''(B(t,s)) \, R(t,s) \, (B(t,s) + B(t,t') - B(s,t')) \,. \qquad (2.4)$$

Note that one can use $B(t, t')$ ($t > t'$) rather than $C(t, t')$ since the Eqs. (2.1),(2.3) and (2.4) are a complete description of the problem together with the identities $B(t, t) = 0$, $\forall t$. In these equations there are two external parameters $T$ and $\mu$.

In deriving these equations a mean over random initial condition $\phi(t = 0)$ (with a Gaussian distribution of variance $C(0, 0)$) is implicit. One could certainly obtain the dynamical equations for a particular initial condition $\phi(t = 0) = \phi_o$ or a different distribution of initial conditions and attempt to study their effects in detail. We shall not do so here but we shall restrict our analysis to the study of the above dynamical equations.

The time-dependent energy-density[12] is

$$\mathcal{E}(t) = \frac{\mu}{2} C(t, t) - 2 \int_0^t ds \; \mathcal{V}'(B(t, s)) \, R(t, s) \; . \tag{2.5}$$

In order to obtain spherically constrained models one has to, on the one hand, let $\mu$ be a function of time $\mu(t)$ and, on the other hand, determine $\mu(t)$ imposing explicitly the spherical constraint. If $C(t, t)$ is set to $\tilde{q}$, then Eq.(2.3) implies

$$\mu(t) \, \tilde{q} = T + 2 \int_0^t ds \; (\mathcal{V}'(B(t, s)) \, B(t, s) + \mathcal{V}''(B(t, s))) \, R(t, s) \; . \tag{2.6}$$

In particular, if $\tilde{q} = 1$ and $\mathcal{V}$ is given by Eq.(1.7) and we compare with the equation determining the Lagrange multiplier $z(t)$ that enforces the spherical constraint in the $p$-spin spherical spin-glass[10], we obtain

$$T \, z(t) = \mu(t) + \frac{p-1}{2} \int_0^t ds \, (C(t, s))^{p-2} \, R(t, s) \; . \tag{2.7}$$

The dynamics is described by the set (2.1)-(2.4) of non-linear coupled integro-differential equations that admit a unique solution for each 'initial auto-correlation' $C(0, 0)$. The dynamical set of equations is *causal*. One can then use a numerical algorithm to iterate the equations and construct the solution step by step in time. A numerical analysis of this type of the model (1.1)-(1.5) with $\gamma < 1$ has been carried out in Ref.[12].

## III. STATIONARY AND NON-STATIONARY DYNAMICS

In the high-temperature phase both time translational invariance (TTI) and the fluctuation dissipation theorem (FDT) hold. Eqs. (2.1)-(2.4) reduce to a single equation. In Refs.[24,25] the model (1.1)-(1.5) with short and long-range correlation have been studied, respectively. The detailed analysis of the resulting equation shows that there is a critical curve $T_c(\mu)$ below which no solution with these characteristics exists. This marks the end of the high-temperature phase. Similar situations occur in the $p$-spin spherical model[30] and in the Random Sine Gordon model[31].

The analytical study of the mean-field dynamical equations in the low-temperature phase requires the use of certain hypotheses. For a purely relaxational dynamics like a Langevin process that approaches an equilibrium situation one can show that the auto-correlation function is a monotonous function of the time difference $\tau$ and that the response function is a positive function of its argument $\tau$. For a general out of equilibrium process the property of monotonicity is not obvious. However, for the kind of models we study it is a quite natural starting point to assume that the systems continuously drift away. As a consequence, though the auto-correlation function *is not* an exclusive function of the time-difference $\tau$ for all times, it is assumed to increase monotonously when the separation between the two times increases:

$$\frac{\partial B(t,t')}{\partial t} \geq 0 \qquad \frac{\partial B(t,t')}{\partial t'} \leq 0 \ . \tag{III.1}$$

The weak-ergodicity breaking hypothesis[20,10] includes the assumption above and, if the particle moves in an infinite dimensional space and there is a finite mass:

$$\lim_{t \to \infty} B(t,t') = b_o \qquad \forall \ t' \ \text{finite} ,$$
$$\lim_{\tau \to \infty} \lim_{t' \to \infty} B(t = \tau + t', t') = q \tag{III.2}$$

and

$$\lim_{t \to \infty} C(t,t) = \tilde{q} \ . \tag{III.3}$$

If the mass is zero from the start the particle diffuses and, in principle, both $b_o$ and $\tilde{q}$ are non-trivial time-dependent functions of $t$ that tend to infinity when $t \to \infty$.

Hence, in the large-times limit, one can separate the relaxation that takes place for large and similar times such that $(t-t')/t' \ll 1$ from the large and widely separated times relaxation such that $(t-t')/t' = O(1)$. The nature of the relaxations in these regimes is clearly different. The former is given by an 'equilibrium' or stationary dynamics while the latter is given by a 'non-equilibrium' or non-stationary dynamics with its maybe associated aging effects[10–12]. This hypothesis has been verified for different models with simulations, numerical analysis and exact results. For instance, the Monte Carlo simulations of various finite dimensional and mean-field spin-glass models[32], the numerical study of the model defined by (1.1) for long-range correlations[12], and the exact analytical results[13] for the model defined by Eqs. (1.6)-(1.7) when $p = 2$ provide evidence for this scenario.

In the large $t'$ limit we thus separate the $t$ dependence in two regimes: for small time-differences $(t-t')/t' \ll 1$ there is equilibrium dynamics obeying the FDT and TTI:

$$\lim_{t' \to \infty} R(\tau + t', t') = r_F(\tau) \ , \quad \lim_{t' \to \infty} B(\tau + t', t') = b_F(\tau) \ , \quad \lim_{t' \to \infty} C(\tau + t', t') = c_F(\tau) \ . \quad \text{(III.4)}$$

The response and displacement functions are related by the FDT theorem

$$r_F(\tau) = \frac{1}{2T} \frac{\partial b_F(\tau)}{\partial \tau} \theta(\tau) = -\frac{1}{T} \frac{\partial c_F(\tau)}{\partial \tau} \theta(\tau) \ . \quad \text{(III.5)}$$

For large time-differences $(t-t')/t' = O(1)$ there is slow dynamics violating both the FDT and TTI:

$$R(t,t') = r(t,t') \ , \quad B(t,t') = b(t,t') \ , \quad C(t,t') = c(t,t') \ . \quad \text{(III.6)}$$

The limiting values for $b$ are then

$$\begin{aligned} b_F(t,t) &= 0 & \lim_{\tau \to \infty} b_F(\tau) &= q \\ b(t,t_-) &= q & b(t,0_+) &= b_0 \end{aligned} \quad \text{(III.7)}$$

(see Eqs.(III.2)-(III.3)). We denote the initial time of the asymptotic regime $s = 0_+$. It has to be understood as a very large time, as opposed to $s = 0$ which is the actual initial time and hence finite. One thus has $b(t,0) \geq b(t,0_+) = b_o$. For fixed $t$ the aging regime extends from $t' = 0_+$ to $t' = t_-$ and FDT regime from $t' = t_-$ to $t' = t$. We shall use $\lim_{t \to \infty} t_-/t = 1$. Calling $q_0$, $q_1$ and $\tilde{q}$ the large time limiting values of $C$

$$\begin{aligned} C(t,t) &= c_F(0) = \tilde{q} & \lim_{\tau \to \infty} c_F(\tau) &= q_1 , \\ c(t,t_-) &= q_1 , & c(t,0_+) &= q_o , \end{aligned} \qquad (\text{III.8})$$

and it follows from these definitions that

$$q = 2\left(\tilde{q} - q_1\right) \qquad b_o = 2\left(\tilde{q} - q_o\right) . \qquad (\text{III.9})$$

Note that the corresponding definitions in the statics are as follows[5,17]. One defines $q_{ab} = \frac{1}{N}\langle \phi_a \phi_b \rangle$ where $a, b$ are replica indices. The equal time correlation $\tilde{q}^{stat}$ corresponds to $q_{aa}$. Within a replica symmetry breaking solution $q_{ab}$ for $a \neq b$ is parametrized by $q(u)$, $0 < u < 1$. While the precise shape of $q(u)$ depends on the model, one has generally a constant $q(u) = q_1^{stat}$ from $u_c < u < 1$ and $q(u) = q_0^{stat}$.

The second and crucial assumption can be called weak long-term memory[10]. It states that the response to any small perturbation applied during a finite interval $[0, t_w]$ eventually decays to zero for a large enough subsequent time $t$. In other words, the integral of the response function $R(t, t')$ over $t'$ in interval $t' \in [t_1, t_2]$ with $t_2 - t_1$ finite vanishes for long enough $t$. In a spin system this means that the thermo-remanent magnetisation decays to zero for a long enough time after having switched-off the (small) magnetic field.

This separation in stationary and non-stationary dynamics allows us to write the dynamical equations (2.1)-(2.4) for the two regimes, separating cleanly each contribution. In what follows we shall refer to the former time-regime $((t - t')/t' \ll 1)$ as the FDT-regime and to the latter time-regime $((t - t')/t' = O(1))$ as the aging regime.

In Section V we shall present some numerical evidence for the separation in these two distinct time-regimes.

The FDT regime

The equations obeyed by the FDT part $((t-t')/t' \ll 1)$ were obtained in Ref.[12]. We recall and discuss them here for completeness.

$$\frac{db_F(\tau)}{d\tau} = 2T + (-\mu + M)\ b_F(\tau)$$
$$-4\int_0^\tau d\tau'\ \mathcal{V}''(b_F(\tau - \tau'))\ r_F(\tau')\ b_F(\tau')$$
$$-4\int_0^\infty d\tau'\ [\ (\mathcal{V}''(b_F(\tau + \tau'))\ r_F(\tau + \tau') - \mathcal{V}''(b_F(\tau')))\ r_F(\tau'))\ b_F(\tau')$$
$$+\ (\mathcal{V}'(b_F(\tau + \tau')) - \mathcal{V}'(b_F(\tau')))\ r_F(\tau')\ ]\ , \tag{III.10}$$

$$\frac{dr_F(\tau)}{d\tau} = (-\mu + M)\ r_F(\tau) - 4\int_0^\tau d\tau'\ \mathcal{V}''(b_F(\tau - \tau'))\ r_F(\tau - \tau')\ r_F(\tau')\ , \tag{III.11}$$

$$c_F(0) = \frac{1}{\mu - \overline{M}} \times$$
$$\left(T + 2\int_0^\infty d\tau'\ \mathcal{V}''(b_F(\tau'))\ r_F(\tau')\ b_F(\tau') + 2\int_0^\infty d\tau'\ \mathcal{V}'(b_F(\tau'))\ r_F(\tau')\right)\ , \tag{III.12}$$

with

$$M \equiv 4\lim_{t \to \infty} \int_0^t ds\ \mathcal{V}''(B(t,s))\ R(t,s)\ . \tag{III.13}$$

$$M_F \equiv 4\int_0^\infty d\tau'\ \mathcal{V}''(b_F(\tau'))\ r_F(\tau')\ , \tag{III.14}$$

$$\overline{M} = M - M_F = 4\lim_{t \to \infty} \int_0^t ds\ \mathcal{V}''(b(t,s))\ r(t,s)\ , \tag{III.15}$$

It is easy to prove that thanks to the FDT relation (III.5) the first two equations collapse and yield only one equation involving $b_F$. However, it also contains contributions from the aging regime through a single, *a priori* unknown quantity 'the anomaly' $\overline{M}$. We shall see that conversely the equations written in the aging regime do not include the details of the FDT solution. Thus, in order to find the full solution for the FDT regime the correct procedure is to first obtain the solution for the aging regime, compute the anomaly, insert it back in the FDT equation, and then solve for the FDT regime.

The equation for the FDT regime is formally identical to the equation one finds studying the equilibrium dynamics à la Sompolinsky though the meaning of the anomaly is different in the two contexts. In the out of equilibrium situation the anomaly represents the memory of the system and, as can be clearly seen from its definition (III.15), it is associated to an integration from the initial time of the asymptotic time-regime up to the final time $t \to \infty$.

In the equilibrium dynamics instead, the initial time is chosen to be $-\infty$ in such a way to propose that the system is at some equilibrium state at a finite time and that it then is able to visit different ergodic components at diverging (with $N$) time scales. The anomalous term is then the contribution from all times starting at $-\infty$ and it is related to contributions from barrier crossing.

Despite the different interpretation for the anomaly one can still borrow or at least formally compare some of the results concerning this equation from the previous works in which the equilibrium dynamics has been studied. The equilibrium analysis has been done in great detail in Refs.[24,25] for the short and long-range model (1.1) and in Ref.[31] for the RSGM. In the high temperature phase one finds $\overline{M} = 0$. However, for $T < T_c(\mu)$ the FDT equation has *no solution* for arbitrary large time differences if one keeps $\overline{M} = 0$. There is a critical time-difference scale beyond which the FDT solution becomes unstable. It is then assumed that the equilibrium dynamics is 'marginal' meaning that the anomaly is chosen to have the minimal value required to remove the instability of the FDT solution.[21,24,25]

Here instead we shall first solve for the well-defined aging solution, compute the anomaly and then return to the study of the asymptotic (large time-difference $\tau$, *i.e.* $\lim_{\tau \to \infty} \lim_{\tau/t' \to 0}$) of the FDT regime. It turns out that, for the models studied here, the value of $\overline{M}$ obtained by this method coincides with the one obtained from the postulate of 'marginal stability'. Thus the detailed behaviour in the FDT regime, such as the power law decay of displacement to the asymptotic value $q$ also coincides.

The aging regime

In Appendix B we derive the set of equations for large and widely separate times, $t > t' \gg 1$ and $(t - t')/t' = O(1)$. The self-consistent procedure is as follows. In this slowly varying time-region, the time-derivatives in the l.h.s. of Eqs.(2.1)-(2.4) are assumed to be small compared to the r.h.s. in the equations and they are then neglected. This is related to the weak ergodicity breaking hypothesis that states that, as time elapses, the relaxation of the systems is slowed down[20,10]. Typically, the integrals in the r.h.s. go from $s = 0$ to $s = t$ or $s = t'$. In order to approximate them we start by separating the contribution of finite times, namely integrals going from the actual initial time $s = 0$ to the starting time of the large time regime $s = 0_+$. These are assumed to be small and hence neglected. This assumption is related to the hypothesis of weak long-term memory[10] and it allows us to write the long-time equations exclusively in terms of the long-time functions $b$ and $r$. Finally we separate the large time-integration intervals to distinguish the FDT and aging contributions. We are typically left with integrals of products of slowly and rapidly varying functions. We approximate then the integrals by replacing the slowly varying functions over the whole integration interval is just constant an equal to the value at the border.

The large-times dynamical equation for the response function $r$ reads

$$0 = r(t,t') \left( -\mu + 4 \int_0^t ds \; \mathcal{V}''(b(t,s)) \, r(t,s) - \frac{2q}{T} \mathcal{V}''(b(t,t')) \right)$$
$$-4 \int_{t'}^t ds \; \mathcal{V}''(b(t,s)) \, r(t,s) \, r(s,t') \qquad (\text{III.16})$$

and that for the displacement correlation function $b$

$$0 = b(t,t') \left( -\frac{\mu}{2} + 2 \int_0^t ds \; \mathcal{V}''(b(t,s)) \, r(t,s) \right) + T + \frac{q}{T}(\mathcal{V}'(q) - \mathcal{V}'(b(t,t')))$$
$$+2 \int_0^t ds \; \mathcal{V}'(b(t,s)) \, r(t,s) - 2 \int_0^{t'} ds \; \mathcal{V}'(b(t,s)) \, r(t',s)$$
$$+2 \int_0^t ds \; \mathcal{V}''(b(t,s)) \, r(t,s) \, b(t,s) - 2 \int_0^{t'} ds \; \mathcal{V}''(b(t,s)) \, r(t,s) \, b(t',s)$$
$$-2 \int_{t'}^t ds \; \mathcal{V}''(b(t,s)) \, r(t,s) \, b(s,t') \; . \qquad (\text{III.17})$$

(NB. We have dropped all the subindices $-, +$ in the limits of the integrals. In particular, we have dropped the $+$ subindex in the zero limits. The zeroes and all the (time) integration variables should be interpreted as being in the asymptotic time regimes.) We shall call these two equations the $r$-Eq. and the $b$-Eq., respectively.

The equation for $C(t,t)$ gives, at large $t$, another equation which contains contributions from the slowly varying parts as well as the FDT parts. In this large times limit, $C(t,t)$ is assumed to have reached its asymptotic value $\tilde{q}$. Thus,

$$0 = -\mu \tilde{q} + T + \frac{q}{T}\mathcal{V}'(q) + 2\int_0^t ds\ \mathcal{V}'(b(t,s))\ r(t,s) + 2\int_0^t ds\ \mathcal{V}''(b(t,s))\ r(t,s)\ b(t,s)\ .$$

(III.18)

For completeness we give the equation for $c(t,t')$. Using similar methods one has:

$$\begin{aligned} 0 =\ & c(t,t')\left(-\mu + 4\int_0^t ds\ \mathcal{V}''(b(t,s))\ r(t,s)\right) + \frac{q}{T}\mathcal{V}'(b(t,t')) \\ & + 2\int_0^{t'} ds\ \mathcal{V}'(b(t,s))\ r(t',s) - 4\int_0^{t'} ds\ \mathcal{V}''(b(t,s))\ r(t,s)c(t',s) \\ & - 4\int_{t'}^t ds\ \mathcal{V}''(b(t,s))\ r(t,s)c(s,t')\ . \end{aligned}$$

(III.19)

We shall call it the $c$-Eq.

We can obtain some relations between the asymptotic values $q,\tilde{q},b_o$ by considering particular values of the times $t,t'$. Letting $t'\to t_-$ in the above equations and using the limiting values (III.7)-(III.9) we find two conditions. Indeed, assuming that the integral in the $r$-Eq. vanishes in this limit, the condition for $r(t,t_-)$ to be non-zero is that the bracket in the following equation vanishes:

$$0 = r(t,t_-)\left(-\mu + 4\int_0^t ds\ \mathcal{V}''(b(t,s))\ r(t,s) - \frac{2q}{T}\mathcal{V}''(q)\right)\ .$$

(III.20)

The solution $r(t,t_-) = 0$ corresponds to the high-temperature phase for which there is no out of equilibrium dynamics. The equation arising from the requirement of the vanishing bracket involves the anomaly: $\overline{M}$ is equal to the integral above when $t\to\infty$ (see Eq.(III.15)). Hence, in order to have a non-trivial low-temperature solution, the anomaly must satisfy

$$\overline{M} = \mu + \frac{2q}{T}\mathcal{V}''(q)\ .$$

(III.21)

This equation is also the marginality condition that determines the anomaly in the equilibrium dynamics treatment. We see here how it arises naturally in the off-equilibirum approach. In addition it is related to the condition of vanishing replicon-eigenvalue in the static - replica - approach (see Section VI).

Similarly, the $b$-Eq. when $t'\to t_-$ yields

$$0 = -\mu + \frac{2T}{q} + 4 \int_0^t ds \; \mathcal{V}''(b(t,s)) \, r(t,s) \tag{III.22}$$

Subtracting Eq. (III.22) from Eq. (III.20) yields

$$q^2 \mathcal{V}''(q) = -T^2 \; . \tag{III.23}$$

This equation determines $q$ as a function of $T$ and the potential correlation $\mathcal{V}$. Interestingly enough the equation is valid for all the potentials for $T < T_c$. It is also independent of the mass and of the dynamical solution of the model and hence $q$ is a purely *geometrical* quantity related only to the potential correlation. Besides, it imposes a condition on $\mathcal{V}''$: it must be negative at $q$ in order to let (III.23) have a sensible solution for $q$. The remaining equation, Eq. (III.20), is important to select the behaviour of the models in the aging regime. Combining both equations one obtains:

$$q = \frac{2T}{\mu - \overline{M}} \tag{III.24}$$

One does not obtain any new equation by letting $t' \to t_-$ in the $c$-Eq. One can check that when combined with Eq. (III.18) it gives back Eq. (III.22), as expected.

Letting $t' \to 0$ in the $b$-eq. one finds:

$$0 = \left(-\frac{\mu}{2} + 2\int_0^t ds \; \mathcal{V}''(b(t,s)) \, r(t,s)\right) b_o + T + \frac{q}{T}(\mathcal{V}'(q) - \mathcal{V}'(b_o))$$
$$+ 2\int_0^t ds \; (\mathcal{V}'(b(t,s)) + \mathcal{V}''(b(t,s)) \, b(t,s)) \, r(t,s) - 2\int_0^t ds \; \mathcal{V}''(b(t,s)) \, r(t,s) \, b(s,0) \; , \tag{III.25}$$

$b_o \equiv b(t,0)$. This is the equation that fixes $b_o$. It is important to note that this equation exists only for a strictly non-zero mass.

Finally, the energy-density (2.5) can be expressed using the separation of FDT and aging regimes as

$$\mathcal{E}(t) = \frac{\mu}{2} \tilde{q} + \frac{1}{T} \left(\mathcal{V}(0) - \mathcal{V}(q)\right) - 2 \int_0^t ds \; \mathcal{V}'(b(t,s)) \, r(t,s) \; . \tag{III.26}$$

## IV. EXTENSIONS OF THE DYNAMICAL THEOREMS TO THE NON-EQUILIBRIUM REGIME

In this Section we review the extensions of the equilibrium theorems for the out of equilibrium dynamics of systems with slow relaxations proposed in Refs.[10,11]. Since we shall mainly use in our calculations the displacement instead of the correlation function we here reexpress these extensions in terms of the displacement function. As opposed to the case of mean-field spin-glass models, in this paper we deal with problems that do not have a normalised correlation function. However, if the mass is non zero, the quadratic potential associated to it ensures that the equal-times correlation function reaches a finite limit at large times. This indeed can be easily checked solving numerically the dynamical equations. Hence, in these cases there is basically no difference between working with the displacement or with the correlation. We shall not discuss here in detail other further extensions associated to the massless cases in which there is unbounded diffusion and the equal-times correlation functions do not reach asymptotically a finite limit[16]. We finally discuss the description of possible singularities at the extremities of the aging regime.

Definitions

One defines the function $\tilde{X}(t,t')$ as:

$$R(t,t') = \tilde{X}(t,t') \frac{\partial B(t,t')}{\partial t'} \qquad (IV.1)$$

with $t \geq t'$. In the FDT regime $\tilde{X}(t,t') = X_F = -1/(2T)$ while in the aging regime it measures the deviation from the FDT theorem.

The mean squared separation $B(t,t')$ monotonically increases when the two times $t$ and $t'$ become more separated. It increases from $B(t,t) = 0$ to its maximum value $B(t,0)$, that explicitly depends on the initial value for the correlation $C(0,0)$: $B(t,0) \equiv C(t,t) + C(0,0) - 2C(t,0)$. Since we shall only consider large times, the maximum value that $B(t,t')$ can take is then $B(t, 0_+) = b(t, 0_+) \equiv b_o$. Note however that, in general, $B(t, 0)$ is different from $b(t, 0_+)$, more precisely, $B(t, 0) \geq b(t, 0_+)$. The monotonicity of $B(t, t')$ allows us to invert the function $B(t, t')$ and to write the function $\tilde{X}$, for instance, in terms of the smaller time and the displacement function, $\tilde{X}(t, t') = X(B(t, t'), t')$. One then assumes that in the large $t'$ limit $X(B(t, t'), t')$ approaches a function that depends only on $B(t, t')$, i.e.

$$X(t, t') \sim X(B(t, t')) \qquad \text{for } t' \gg 1 . \tag{IV.2}$$

We shall also assume that $|X(B)|$ is a monotonously decreasing function of $B$. In Section V we shall present results from the numerical solution of the dynamical equations that supports the assumption made in Eq.(IV.2) and the further assumption of monotonicity. The property of monotonicity permits to invert $X(B)$ to have a function $B(X)$. This function is useful to compare the asymptotic dynamical values with the equilibrium ones and it is somehow related[10] to the - replica - Parisi function $b(u) = 2(\tilde{q}^{stat} - q(u))$.

Similarly for three given times, thanks to the monotonicity of the displacement $B$, one can always write the relation

$$B(t, t') = \tilde{f}(B(t, s), B(s, t'), t') . \tag{IV.3}$$

One then assumes that when all times are large, $t > s > t'$ the explicit time dependence disappears and

$$B(t, t') = f(B(t, s), B(s, t')) \qquad t > s > t' \to \infty . \tag{IV.4}$$

The function $f$ connects any three correlation functions; it has been called a triangle relation. Formally, one can also define the reciprocal function $\overline{f}(B', B)$ such that

$$B(s, t') = \overline{f}(B(t, s), B(t, t')) . \tag{IV.5}$$

Similar and in the end equivalent definitions of $X$ and $f$ could be given in terms of $C(t, t')$.

Properties of the triangle relations

Let us recall the general properties of the function $f(x,y)$. By definition $f(x,y)$ is associative $f(f(x,y),z) = f(x,f(y,z))$. Since $b(t,t')$ is an increasing function of the separation between $t$ and $t'$ one must have

$$f(x,y) \geq \max(x,y) . \tag{IV.6}$$

One defines 'fixed points' $b_i^*$ such that $f(b_i^*, b_i^*) = b_i^*$. Clearly $f(x,x) > x$ for $x$ between two fixed points. One can show that

$$b^* \leq x \qquad f(b^*, x) = f(x, b^*) = x , \tag{IV.7}$$

$$x \leq b^* \qquad f(x, b^*) = f(b^*, x) = b^* . \tag{IV.8}$$

Hence the relation between a fixed point $b^*$ and any other point $x$ is in this case 'maximum'. Between two fixed points $b_1^* < x < b_2^*$, one can show, under the assumption that $f(x,y)$ is *smooth enough*[34], i.e. that it has a formal series expansion, and using the fact that there exists an identity ($b_1^*$) and a zero ($b_2^*$):

$$\begin{aligned} i. & \quad f(x,y) = f(y,x) & f \text{ is commutative} \\ ii. & \quad f(x,y) = \jmath^{-1}(\jmath(x)\jmath(y)) & f \text{ is isomorphic to the product} \end{aligned} \tag{IV.9}$$

with $\jmath(b_1^*) = 1$ and $\jmath(b_2^*) = 0$. This implies $y = \overline{f}(x,z) = \jmath^{-1}\left(\jmath(z)/\jmath(x)\right)$, $z \geq x$.

It is useful to explicitly compute some derivatives of $f(x,y)$ and $\overline{f}(x,z)$ w.r.t. $x$, $y$ and $z$ when they are isomorphic to the product and division. For $x, y \in (b_1^*, b_2^*)$ one finds

$$-\left.\frac{\partial \overline{f}(x,y)}{\partial x}\right|_{x=y \to b_1^*} = \left.\frac{\partial \overline{f}(x,y)}{\partial y}\right|_{x=y \to b_1^*} = 1 . \tag{IV.10}$$

Moreover, since $f(b^*, x) = x$ and $\overline{f}(b^*, x) = x$ for all $x > b^*$ then $\partial f(b^*, x)/\partial x = 1$ and $\partial \overline{f}(b^*, x)/\partial x = 1$ for all $x > b^*$. Instead, $f(b^*, x) = b^*$ and $\overline{f}(b^*, x) = b^*$ for all $x < b^*$, then $\partial f(b^*, x)/\partial x = 0$ and $\partial \overline{f}(b^*, x)/\partial x = 0$ for all $x < b^*$. Thus, $\partial f(b^*, x)/\partial x = \theta(x - b^*)$.

It is useful to note that the form (IV.9-ii) can be expressed as: the form:

$$B(t,t') = \jmath^{-1}\left(\frac{h(t)}{h(t')}\right) \quad h(t) = B(t,t_1) \tag{IV.11}$$

where $t_1$ is an arbitrary fixed time such that $t_1 < t' < t$.

However, the requirement that $f$ has a formal series expansion[34] turns out to be too restrictive and for certain dynamical problems the large time dynamical equation may not admit a solution with an analytic $\jmath$ or $\jmath^{-1}$ in the whole interval $(b_1^*, b_2^*)$.

Indeed, studying the $p=2$-spherical spin-glass model[13] one finds that the *exact* solution to the dynamical equations can be written, in the regime of large and widely separated times, as a function $f$ isomorphic to the product as in Eq.(IV.9), but with a not well-behaved $\jmath(\lambda)$ when $\lambda \to 1$. This is indeed related to the fact that the time derivatives $\partial c(t,t')/\partial t = \partial c(t,t')/\partial t' = 0$ when $t' \to t_-$ that implies that the inverse is not defined in this limit.

More precisely, the zero temperature solution corresponds to[13]

$$\jmath^{-1}(\lambda) = 2\sqrt{2}\,\frac{\lambda^{3/4}}{(1+\lambda)^{3/2}} \tag{IV.12}$$

with $\lambda = h(t')/h(t) = t'/t$. $\jmath^{-1}$ has a vanishing derivative when $\lambda \to 1$ and hence $\jmath'(\lambda)|_{\lambda \to 1} \to \infty$. $\lambda \to 1$ for $\jmath^{-1}$ corresponds to $b' \to b$ for $f$ and $\overline{f}$ or $t' \to t_-$ for $c$ and $b$; *i.e.* it is the beginning of the aging regime.

One of the main result of the present paper is that many of the models studied here do not admit a smooth solution when $b' \to b$ but, instead, are solved in the aging regime by an ansatz still of the form (IV.9) with a non-analytic $\jmath^{-1}$ at $\lambda \to 1$.

General organisation of fixed points

We concentrate on the analysis of the long-time dynamics for which the functions defined above have been proposed. Having neglected the time-dependence in the function $X$ and $f$ suggests that the evolution of the system should be measured in terms of the displacement-value (or the correlation-value) instead of in terms of the times. This implies that time-scales are replaced by correlation-scales. A correlation scale is the range of correlations between two fixed points of the function $f$. We call a 'blob' a discrete correlation scale between two separate fixed points, for which the function $f$ is proposed to satisfy Eq. (IV.9).

It is clear that $B(t,t) = 0$ is a fixed point of $f$ for all the models we consider. The separation of large times into close and separated amounts to assuming the existence of a fixed point at $B(t,t_-) = q$ that marks the end of the FDT correlation scale and the begining of the out of equilibrium correlation scales, the aging regime.

When the times are close to each other, the displacements are homogeneous functions of time, $B(t,t') = B(t-t')$. Calling $\tau_1 = t - t'$, $\tau_2 = t - t''$, $\tau_3 = t'' - t'$ and $B_1 = B(\tau_1)$, $B_2 = B(\tau_2)$, $B_3 = B(\tau_3)$ one can immediately show that a function $f$ as defined above exists in the FDT regime. Indeed, the displacement functions can be inverted to give $\tau_2 = \tau_2(B_2)$, $\tau_3 = \tau_3(B_3)$ and then using $\tau_1 = \tau_2 + \tau_3$ we have $B_1 = B(\tau_1) = B(\tau_2 + \tau_3) = B(\tau_2(B_2) + \tau_3(B_3))$. Thus one finds that in the FDT regime $B(\tau) = \jmath^{-1}(\exp(\tau))$. As regards the function $X$ it is just the constant associated to the FDT theorem in this correlation scale, $X = -1/(2T)$.

Another particular value of the displacement function is $b_o$. Since it is defined as the maximum value the displacement function can take in the large-times dynamics (for $\mu > 0$), it must be necessarily a fixed point of $f$. The displacement cannot go beyond this value. Moreover $b(t, 0_+) = b(t', 0_+) = b_o$.

The strategy is to solve the large-times dynamical equations, $viz.$ the $r$-Eq., $b$-Eq. and $c$-Eq, (Eqs.(III.16)-(III.17)) using the general properties of the functions $X$ and $f$. The dynamical equations for each particular model, $i.e.$ for each particular $\mathcal{V}(x)$, will determine the form of $X$ and $f$.

## V. REPARAMETRISATION INVARIANT EQUATIONS FOR THE AGING REGIME

In this Section we write the dynamical equations for the aging regime, Eqs.(III.16)-(III.19) in terms of the functions $X$ and $f$ defined and decribed in the previous section. We discuss the implications of those assumptions and we also present numerical evidence for them. We finally describe the strategy we shall follow in the two next Sections to find an analytic solution to the non-stationary dynamics.

Neglecting the time-derivatives in the dynamical equations and the explicit dependence on time of the functions $X$ and $f$ amounts to introducing an artificial time-reparametrisation invariance into the large-times dynamical equations. The solutions we shall find are as a consequence invariant under time reparametrisations. This problem has already been encountered when solving the asymptotics of mean-field spin-glass models. The question on how to select the unique actual solution from the time-reparametrisation invariant family of solutions has not been answered yet. Indeed, this is already a well-know problem in the theory of non-linear one time differential equations. It is sometimes called the *matching problem* and it has been solved only for some particular cases. How to match the solutions we find for short time differences with one representative of the family of large time-differences solutions remains as a hard open problem.

We can thus completely eliminate the times in the $r$-Eq., $b$-Eq. and $c$-Eq, and rewrite them just in terms of the displacement $b$. With this aim we define

$$F[b] \equiv -\int_b^q db' \, X(b') \qquad \overline{M}[b] \equiv 4\int_b^q db' \, \mathcal{V}''(b')X(b') \ . \tag{V.1}$$

The function $\overline{M}[b]$ is related to the anomaly: $\overline{M} = \overline{M}[b_o]$, see Eq. (III.15). These functions could be also defined in such a way to contain the FDT regime. The upper limit in the integrals should then be set to $b' = 0$. In this case the second integral would be related to $M$, Eq.(III.13).

We can then rewrite the $r$-Eq., Eq.(III.16), as

$$0 = \frac{\partial}{\partial t'} \left( (-\mu + \overline{M}) F[b(t,t')] + \frac{q}{2T} \overline{M}[b(t,t')] \right)$$
$$-4\mathcal{V}''(b(t,t')) \, r(t,t') \, F[q] - 4\frac{\partial}{\partial t'} \int_{b(t,t')}^{q} db' \, \mathcal{V}''(b') \, X[b'] \, F[\overline{f}(b', b(t,t'))] \, . \qquad (\text{V.2})$$

Using $F[q] = 0$ and integrating over $t'$ one finds:

$$0 = F[b] \left( -\mu + \overline{M} \right) + \frac{q}{2T} \overline{M}[b] + 4 \int_q^b db' \, \mathcal{V}''(b') \, X[b'] \, F[\overline{f}(b', b)] \qquad (\text{V.3})$$

the integration constant being zero as can be seen by evaluating the resulting equation at $b = q$. This equation is in fact equivalent to a first integral of the $r$-Eq.

The $b$-Eq., Eq.(III.17), can be rewritten as:

$$0 = \left( -\mu + \overline{M} \right) b + 2T + \frac{2q}{T} \left( \mathcal{V}'(q) - \mathcal{V}'(b) \right) + 4\mathcal{V}'(b_0) F[b'_0] + 4 \int_{b_0}^{q} db' \, \mathcal{V}'(b') \, X[b'] +$$
$$4 \int_{b_0}^{b} db' \, \mathcal{V}''(b') \, F[\overline{f}(b,b')] + 4 \int_{b_0}^{q} db' \, \mathcal{V}''(b') \, b' \, X[b'] - 4 \int_{b_0}^{b} db' \, \mathcal{V}''(b') \, \overline{f}(b,b') \, X[b']$$
$$+4 \int_q^b db' \, \mathcal{V}''(b') \, \overline{f}(b',b) \, X[b'] \, , \qquad (\text{V.4})$$

where we denote $b'_0 = b(t', 0)$. For $t > t' > 0$, $b(t, 0) \geq b(t', 0)$. Since in the large time $t'$ limit $b(t', 0) = b'_o$ must be a fixed point, for all $t' > 0$: $b_o = b(t, 0) = f(b(t, t'), b(t', 0)) = b(t', 0) = b'_o$.

Both equations (V.3) and (V.4) are evidently time-reparametrisation invariant. In the same way we can re-write the $c$-Eq., Eq.(III.19), in terms of $b$.

The $b$-Eq. (V.4) evaluated in $b \to q$ and $b \to b_o$ gives back Eqs. (III.22) and (III.25), respectively, that in terms of the functional $F$ and $\overline{M}$ read

$$0 = (-\mu + \overline{M}) q + 2T \, , \qquad (\text{V.5})$$
$$0 = -\mu b_o + 2T + \frac{2q}{T} \left( \mathcal{V}'(q) - \mathcal{V}'(b_o) \right) + 4\mathcal{V}'(b_o) F[b_o] + 4 \int_{b_o}^{q} db' \, \frac{\partial}{\partial b'} \left( \mathcal{V}'(b') \, b' \right) X[b'] \, , \qquad (\text{V.6})$$

again, the latter exists only if $\mu \neq 0$ from the start.

The solution to the aging regime amounts, in this language, to solving Eqs.(V.3) and (V.4). Our strategy is as follows:

A. We propose the existence of only one discrete correlation scale (one blob) with a constant

$X[b] = x$. We propose a completely regular $\overline{f}$; by this we mean that the integration of a finite function of $b'$ times $\overline{f}(b',b)$ between $b$ and $q$ tends to zero when $b \to q$. We find that only the $p$-spin spherical spin-glass and the limit of vanishing mass of the RSGM are solved by an ansatz of this type.

B. We propose the existence of only one discrete scale but we allow for non-analytic solutions around $b \sim q$. This implies, in particular, that the integral above contributes in the limit $b \to q$. We find the conditions the potential correlation must satisfy in order to allow for such a solution. Model (1.1) with short-range correlations and the massive RSGM are solved in this way.

C. We propose an ultrametric ansatz for all the correlations in the aging regime, namely $\overline{f}(b',b) = \max(b',b)$, $b',b > q$. We show the condition the potential correlation must satisfy to admit this ansatz as a solution. Model (1.1) with long-range correlations belongs to the family of models with an ultrametric dynamical solution. We present this analysis in the Section VII.

In principle it is posible to show that the dynamical equations do not admit another type of solution with, for example, two discrete scales (two blobs) and different constant $x$ inside each of the blobs or many other possibilities. This is a rather painful task that we do not do in this paper. We just study the three cases we describe above.

### A. Numerical evidence for the two regimes

The numerical solution of the full set of dynamical demonstrates the existence of the two separated time-regimes and also the existence of a function $X(B)$ for large times. One convenient way to show this is by plotting the equivalent of the thermo-remanent magnetisation of a magnetic system

$$m(t,t') \equiv \int_0^{t'} dt''\, R(t,t'') \qquad\qquad \text{(V.A.1)}$$

vs the displacement function $B(t,t')$. If $R(t,t') = X(B(t,t'))\, \partial B(t,t')/\partial t'$ for large $t'$, then $X(B)$ is the local slope of these curves.

In Fig. 1 and 2 we plot $m(t,t')$ vs. $B(t,t')$ for the power law model with long $\gamma = .5$ and short $\gamma = 1.5$ correlations, respectively. The external paramters are $\mu = 0.2$, $\theta = 2.5$ and $T = 0.2$ and we choose $C(0,0) = 0$. Three curves for three times $t$ are included in both figures: $t = 200, 300, 400$ (1000, 1500 and 2000 iterations of the algorithm with a step $h = 0.2$; all our numerical results have been performed with $h = 0.2$). They show clearly two $B$ regimes: for $B \leq q$ the slope is constant and equal to $-1/(2T) = -0.25$ (FDT holds). For $B > q$ and if the potential correlation is long-range the slope is not constant while if it is short-range the slope is still compatible with a constant $|x| < |-1/(2T)|$. The breaking point $q$ and end point $b_o$ of the displacement can be read from the figures and as we shall see in Sections VI and VII they coincide nicely with the analytical predictions.

The curves for different total times $t$ do not exactly lie upon each other showing that the numerical $X(B)$ still depends on $t$. We expect that for sufficiently large times they will converge to the analytical prediction that we describe in the Sections VI and VII and that we include in these figures as the broken lines. We can see that the numerical convergence is indeed rather slow. In the insets we plot the same curves for the same choice of parameters but for the high-temperature phases, $T = 1$. $t = 400$. In both figures the slope is $-.5$ for all $B$.

FIGURES

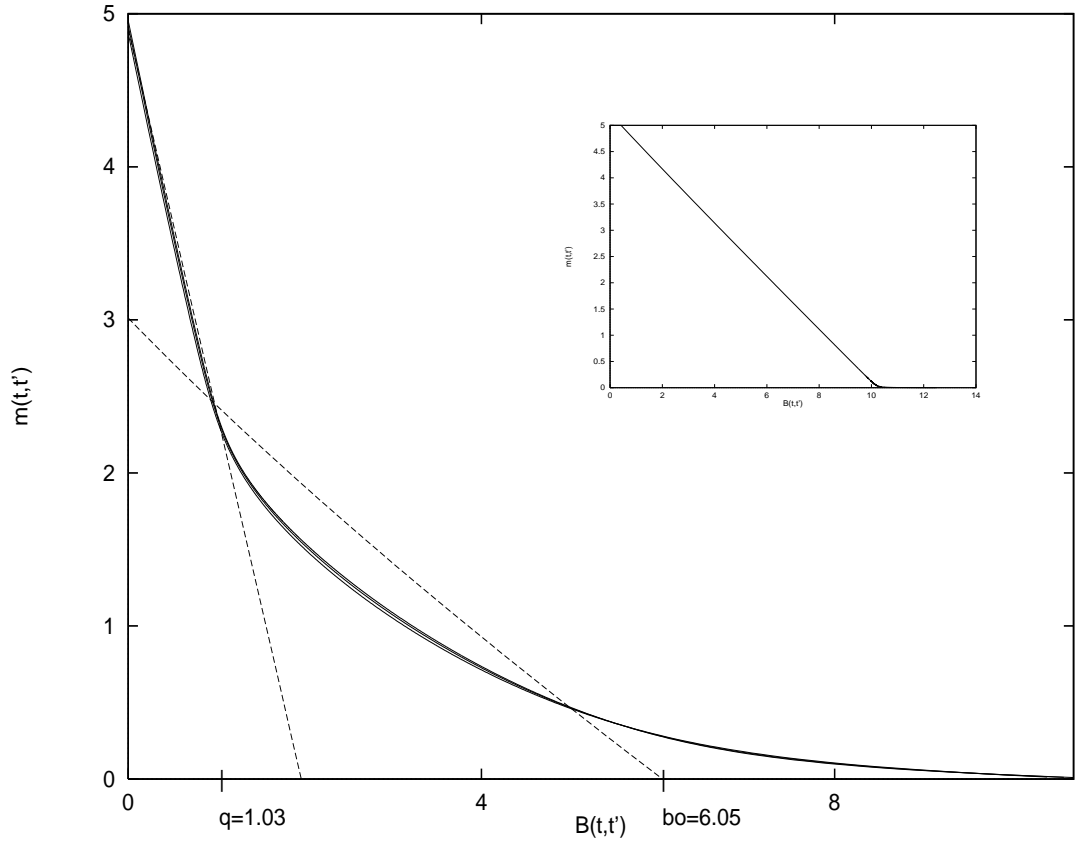

FIG. 1. $m(t,t')$ vs. $B(t,t')$ for the long-range power law model. $\gamma = .5$, $\mu = 0.2$, $\theta = 2.5$. For the full lines $t = 200, 300, 400$. In the main plot $T = 0.2$ while in the inset $T = 1. > T_c$. Note that with the choice $C(0,0) = 0$ one has automatically $C(t,0) = 0$ and thus $B(t,0) = C(t,t)$.

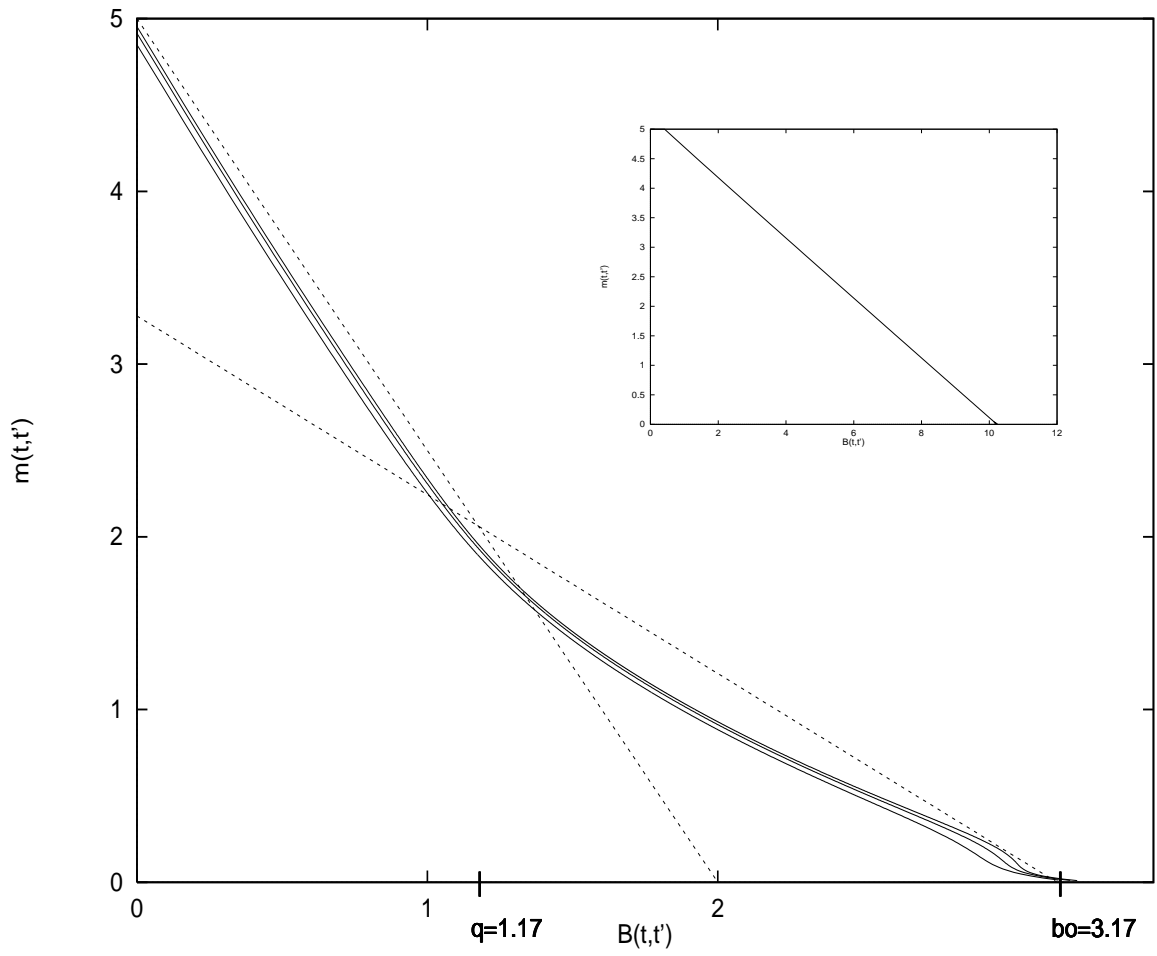

FIG. 2. $m(t,t')$ vs. $B(t,t')$ for the short-range power law model. $\gamma = 1.5$, $\mu = 0.2$, $\theta = 2.5$. For the full lines $t = 200, 300, 400$. $T = 0.2$ while in the inset $T = 1. > T_c$. $q \sim 1.17$, $b_o \sim 3.17$.

# VI. SOLUTIONS TO THE DYNAMICAL EQUATIONS: SHORT-RANGE POTENTIALS

In this Section we apply a one blob ansatz to the aging regime equations and we show that those associated to the short-range potentials are solved in this way.

## A. Asymptotic values: comparison with the statics

We start by studying the asymptotic values $q$, $b_o$, and $x$. Though a deep understanding of the relation between the dynamical $X[b]$ and the statical replica function $x[q]$ is still missing it is by now clear that, at least formally, they play similar roles. We compare here the aymptotic values $q$ and $b_o$, and the parameter $x = X[b]$ inside the blob, to the static[17] $q^{stat}$, $b_o^{stat}$ and $x^{stat}$.

Equations (III.22) and (III.25) with $X[b] = x$ are identical to the ones obtained within the replica analysis[17] as the extremum conditions for the static $q$ and $b_o$. This is a consequence of an algebraic relation between dynamical and static (replica) approaches[35]. Instead, the third equation, Eq.(III.21), dynamical in origin, is formally equal to the condition to have a vanishing replicon-eigenvalue of the replica treatment. It does not coincide with the third equation arising from the replica recipe, *i.e.* the one obtained maximising the free-energy w.r.t. $x$. As a consequence, the dynamical value for $x$ does not coincide with the static one. This implies that also the values for $q$ and $b_o$ differ since $x$ enters in their equations. The dynamical phase diagram turns out to be different from the static, the dynamical transition temperature is higher than the static critical temperature for a range of masses. Moreover, the dynamical asymptotic energy is higher than the equilibrium energy. This discrepancy has already been observed for the special case of the *p*-spin spherical spin-glass and it seems to be related to a peculiar organisation of the metastable states of the TAP[37] free-energy landscape, with the existence of a threshold (higher than the equilibrium free-energy, below which infinite barriers appear[10,38]). A more detailed analysis along these lines is in order to determine if this is at the origin of the difference between static and dynamical asymptotic values in these models too.

If the model is massive, the particle is constrained to move in a restricted region of the space and hence $b_o$ and $\tilde{q}$ are finite. If we take the limit $\mu \to 0$ afterwards we shall see that $b_o \to \infty$ in such a way that $\mu b_o$ remains finite. Instead the case of a strictly massless model is more delicate. Let us now obtain the equations that fix $q$, $b_o$ and $x$ for a massive model. Eq. (III.23) determines $q$. When $x$ is constant the equations that fix $b_o$ and $x$ follow from Eqs. (V.5) and (V.6) and they read

$$\frac{\mathcal{V}'(q) - \mathcal{V}'(b_o)}{b_o - q} = \frac{\mu T}{2q} , \qquad (\text{VI.A.1})$$

$$b_o - q = -\frac{1}{2Tx}\left(\frac{2T}{\mu} - q\right) , \qquad (\text{VI.A.2})$$

while Eq.(VI.A.3) gives $\tilde{q}$:

$$\tilde{q} = \frac{1}{\mu}\left[T + \frac{q}{T}\mathcal{V}'(q) + 2x\left(\mathcal{V}(q) - \mathcal{V}(b_o)\right) + 2x(q\mathcal{V}'(q) - b_o\mathcal{V}'(b_o))\right] . \qquad (\text{VI.A.3})$$

The anomaly $\overline{M}$ can be immediately computed:

$$\overline{M} = 4x\left(\mathcal{V}'(q) - \mathcal{V}'(b_o)\right) . \qquad (\text{VI.A.4})$$

and can be used to solve the FDT equations (see Section III).

The asymptotic energy-density reads

$$\mathcal{E}(t) = \frac{\mu}{2}\tilde{q} + \frac{1}{T}\left(\mathcal{V}(0) - \mathcal{V}(q)\right) - 2x\left(\mathcal{V}(q) - \mathcal{V}(b_o)\right) \qquad (\text{VI.A.5})$$

and thus it is determined by the asymptotic values $q, b_o$ and $x$. Since $x$ does not coincide with the static - replica - the asympotic dynamical energy differs from the static one. We shall show below some numerical results on this point (Fig. 7).

## B. The $\mu - T$ phase diagram

In Fig.3 we present the phase diagram for the power law model with $\gamma = 1.5$ and $\theta = 2.5$.

The equation for $q$, Eq.(III.23), determines a first critical temperature (independent of the mass) given by

$$T_c^{(1)} = \sqrt{-(q^*)^2 \mathcal{V}''(q^*)} \,, \qquad \text{(VI.B.1)}$$

with $q^*$ the value of $q$ at which $q^2 \mathcal{V}''(q)$ attains its maximum, i.e.

$$\frac{q^* \mathcal{V}'''(q^*)}{2 \mathcal{V}''(q^*)} = -1 \,. \qquad \text{(VI.B.2)}$$

The glass phase cannot extend beyond $T = T_c^{(1)}$ since there is then no solution to Eq.(III.23).

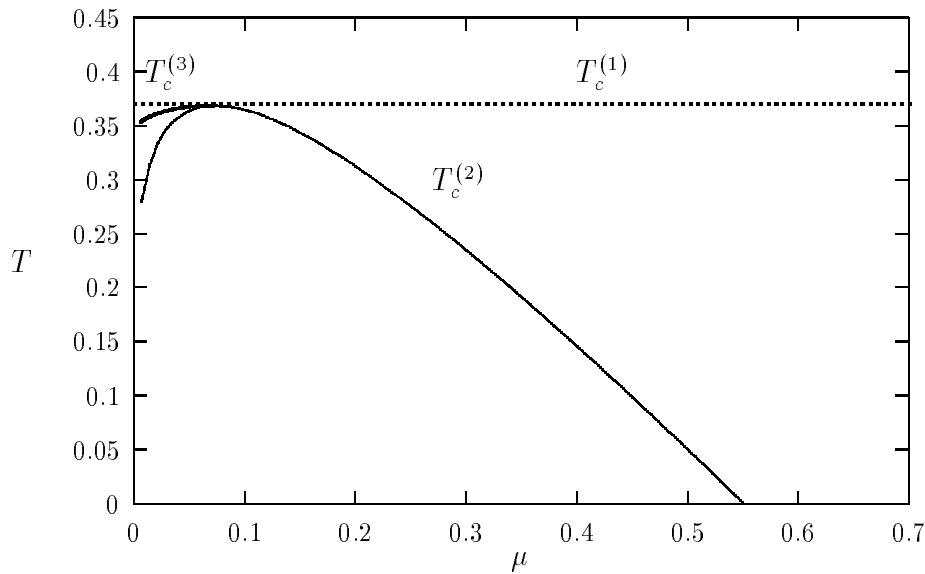

FIG. 3. Phase diagram for the power law model with $\gamma = 1.5$ and $\theta = 2.5$.

Just from their definitions $b_o \geq q$. This condition determines a critical line $T_c^{(2)}(\mu)$. Eq.(VI.A.1) when $b_o \to q$ yields

$$T_c^{(2)}(\mu) = \frac{q(T)\mu}{2} \,. \qquad \text{(VI.B.3)}$$

where $q(T)$ is the solution to Eq.(III.23). The general equation for this line is:

$$\mathcal{V}''\left(\frac{2T_c^{(2)}}{\mu}\right) = -\frac{\mu^2}{4} . \qquad \text{(VI.B.4)}$$

This same equation is found in the replica treatment as the condition to have a a vanishing replicon eigenvalue. Statically this determines the line below which the replica symmetric solution becomes unstable and hence the transition line from the RS solution to the 1RSB solution[17]. Eq.(VI.B.4) delimits a domain in the $\mu$-$T$ plane were the aging solution exists. When $T \to 0$ we find a finite critical $\mu = \mu_c = 2\sqrt{|\mathcal{V}''(0)|}$ beyond which the above equation has no solution. When $\mu \to 0$ and the potential is short-range (i.e. $\mathcal{V}''(x)$ decreases faster than $1/x^2$ at infinity) $T_c^{(2)}(\mu) \to 0$.

For the power law model the explicit formula for $T_c^{(2)}(\mu)$ is

$$T_c^{(2)}(\mu) = \frac{\mu}{2}\left(\left(\frac{2\gamma}{\mu^2}\right)^{1/(\gamma+1)} - \theta\right) . \qquad \text{(VI.B.5)}$$

If $\gamma > 1$, $T_c^{(2)}(\mu \to 0) \to 0$ and for $\gamma = 1$, $T_c^{(2)}(\mu \to 0) = 1/\sqrt{2}$.

Returning to the general potential correlation and evaluating $x$ from Eq.(VI.A.2) at the critical line $b_o = q + \epsilon$, $\epsilon \to 0$, one obtains

$$x = x_c = \frac{q}{4T}\frac{\mathcal{V}'''(q)}{\mathcal{V}''(q)} . \qquad \text{(VI.B.6)}$$

with $q = 2T_c^{(2)}/\mu$ on the critical line. One sees that as long as the function $q^2\mathcal{V}''(q)$ is increasing (small values of $q$, $q < q^*$), one automatically has $|x_c| < 1/(2T)$. Thus, even if the transition is continuous for $q$ and $b_0$ there is a jump in $x$, from the value $x_c$ below the critical line to the high-T phase value $x = x_F = -1/(2T)$.

The value $x_c = -1/(2T)$ is exactly reached at the temperature $T_c^{(2)} = T_c^{(1)}$ where $q$ attains its maximal value. Equations (VI.B.1) and (VI.B.4) determine the tricritical point $(\mu_c^*, T_c^*)$, at which $b_o = q$ and $x = -1/(2T)$ simultaneously. For $\mu < \mu_c^*$ the nature of the transition changes. The phase boundary is now determined by the condition $x_c = -1/(2T)$, while $b_0$ at the transition is no longer equal to $q$. Setting $x = -1/(2T)$ in the equations (VI.A.1) and (VI.A.2) one obtains $b_o = 2T_c^{(3)}/\mu$ at the transition and the following equations for the transition temperature:

$$\mathcal{V}'(q) - \mathcal{V}'\left(\frac{2T_c^{(3)}}{\mu}\right) = \frac{(T_c^{(3)})^2}{q} - \frac{\mu T_c^{(3)}}{2}, \tag{VI.B.7}$$

with $q$ given, as always, by Eq.(III.23). The transition is now continuous in $x$ but discontinuous in $q$. $T_c^{(3)}(\mu)$ is well above the transition line for the statics and it reaches a finite limit at $\mu \to 0$, as opposed to $T_c \to 0$ when $\mu \to 0$ in the statics. For the potential correlation (1.5) one finds that when $\mu \to 0$, $(T_c^{(3)}(\mu))^2 \to \frac{\theta^{1-\gamma}}{2(\gamma-1)}(\frac{\gamma-1}{\gamma})^\gamma$ and $q = \theta/(\gamma-1)$.

For model (1.1) with $\mathcal{V}$ given by Eq.(1.5) and $\gamma > 1$, these conditions imply a phase diagram identical to the one described in Ref.[24] for the equilibrium dynamics. It is not clear why it should be a difference between static and equilibrium dynamic transitions (and furthermore between static and asymptotic equilibrium dynamic energies) especially if one remembers that the initial assumption of the latter approach is that the evolution starts at an equilibrium state. In the out of equilibrium evolution there is *a priori* no reason for the two transition temperature to coincide and indeed one knows this happens in real systems as glasses. We shall come back to this discussion later when we shall present the numerical results for the asymptotic energy-density *vs.* the static one.

### C. Equations within a blob with $X[b] = x$

Using $X[b] = x$:

$$F[b] = -x(q-b) \qquad \overline{M}[b] = 4x\left(\mathcal{V}'(q) - \mathcal{V}'(b)\right), \tag{VI.C.1}$$

and the first integral of the $r$-Eq., Eq. (V.3), and the $b$-Eq., Eq. (V.4), read

$$0 = (b-q)\left[-\mu + 4x\left(\mathcal{V}'(q) - \mathcal{V}'(b_o)\right)\right] + 4q\left(\frac{1}{2T} + x\right)\left(\mathcal{V}'(q) - \mathcal{V}'(b)\right)$$
$$+ 4x\int_q^b db'\, \mathcal{V}''(b')\, \overline{f}(b', b) \tag{VI.C.2}$$

$$0 = b\left[-\mu + 4x\left(\mathcal{V}'(q) - \mathcal{V}'(b_o)\right)\right] + 2T + 4q\left(\frac{1}{2T} + x\right)\left(\mathcal{V}'(q) - \mathcal{V}'(b)\right)$$
$$+ 4x\int_q^b db'\, \mathcal{V}''(b')\, \overline{f}(b', b). \tag{VI.C.3}$$

We see that these equations coincide if we use $\mu - \overline{M} = 2T/q$. We shall concentrate then on the $r$-Eq. (VI.C.2).

### D. Analytic ansatz at $t' \sim t$

We can now show that a regular solution at $t' \to t_-$ cannot exist for all the models we consider. In fact, if $\overline{f}$ is regular at $b \to q$ one can evaluate Eq.(VI.C.2) and its variations with respect to $b$ in this limit. If $\overline{f}$ and all its variations with respect to $b$ are smooth when $b \to q$, one can set the integral in each equation to zero (the limits collapse and the integrands are regular by hypothesis) when evaluating at $b \to q$. However, if this is so, one gets new conditions on $x, q$ and $b_o$ that are not always compatible.

More precisely, the first new equation arises from the second variation of the $r$-Eq. w.r.t. $b$. When $b \to q$ it implies

$$x = \frac{q}{2T} \frac{\mathcal{V}'''(q)}{\mathcal{V}''(q)} \,, \qquad \text{(VI.D.1)}$$

where we have used $\partial \overline{f}(b', b)/\partial b|_{(b' \to b) \to q} = 1$ and we have set the remaining integral to zero. For general models, $i.e.$ general $\mathcal{V}$s, this condition is not compatible with the $x$ arising from Eq.(VI.A.2) above.

In an alternative way, one can investigate the equation the potential correlation $\mathcal{V}$ has to satisfy in order to allow for a regular solution to the dynamical equation. Combining Eqs. (III.23), (VI.A.1) and (VI.A.2) with Eq.(VI.D.1) yields

$$\frac{\mathcal{V}'(q) - \mathcal{V}'(b_o)}{b_o - q} = -\frac{\mathcal{V}''(q)}{1 - (b_o - q) \frac{\mathcal{V}'''(q)}{\mathcal{V}''(q)}} \,. \qquad \text{(VI.D.2)}$$

We can now consider the temperature $T$ as a varying parameter and then $q$ as the independent variable in a differential equation for $\mathcal{V}(q)$. Calling $y(q) \equiv \mathcal{V}'(q)$ and $y_o \equiv \mathcal{V}'(b_o)$ we have:

$$-(y - y_o)(y' - (b_o - q)y'') = (b_o - q) y'^2 \ . \tag{VI.D.3}$$

The potential correlation has to satisfy this equation in order to allow for a regular solution.

We have checked that from the potential correlations described in the Introduction only the one associated to the $p$-spin spherical model with $p \geq 2$ admits a solution with a well-behaved $\jmath^{-1}$ when $\lambda \to 1$ (the case $p = 2$ is particular because the first derivative of $\jmath^{-1}$ is zero). The $x$ arising from the second variation of Eq. (VI.C.2) given by Eq.(VI.D.1) is compatible with the $x$ arising from Eq.(VI.A.2). All the higher variations are trivial and do not give further constraints. Conversely, if one perturbs the original $p$-spin Hamiltonian with an arbitrary small term associated to a different $p$ in the manner of Ref.[36], one immediately sees that a regular solution for the dynamics is not more allowed. One can say that the $p$-spin spherical spin-glass is somehow a marginal model.

The problem for a general $\mathcal{V}$ is in fact so constrained that it is not hard to believe that the potential correlation associated to the $p$-spin spherical spin-glass be the unique function having a regular solution.

### E. Non-analytic solution at $t' \sim t$

Using $\mu - \overline{M} = 2T/q$ the remaining equation (VI.C.2) reads

$$0 = -\frac{T}{q}(b - q) - 2q\left(x + \frac{1}{2T}\right)(\mathcal{V}'(b) - \mathcal{V}'(q)) + 2x \int_q^b db' \, \mathcal{V}''(b') \jmath^{-1}\left(\frac{\jmath(b)}{\jmath(b')}\right) \ . \tag{VI.E.1}$$

It is useful now to remember the original time-dependence inside the integral using

$$b(t, t') = \jmath^{-1}\left(\frac{h(t')}{h(t)}\right) \ . \tag{VI.E.2}$$

Thus, defining a new variable $u$ and a new function $B(u)$ as[10]:

$$\exp(-u) \equiv h(t) \qquad b(t, t') = q + B(u - u') \equiv \jmath^{-1}\left(\frac{h(t')}{h(t)}\right) \ , \tag{VI.E.3}$$

Eq.(VI.E.1) becomes

$$0 = B(u)\mathcal{V}''(q) - \mathcal{V}'(q+B(u)) + \mathcal{V}'(q) - \frac{2xq}{T}\mathcal{V}''(q)\int_0^{B(u)} dB(u')\,\mathcal{V}''(q+B(u'))\,B(u-u')\,,$$

(VI.E.4)

or, integrating by parts the last term,

$$0 = -B(u)\mathcal{V}''(q) + \mathcal{V}'(q+B(u)) - \mathcal{V}'(q)$$
$$+\frac{2xq}{T}\mathcal{V}''(q)\left[-\mathcal{V}'(q)B(u) + \frac{\partial}{\partial u}\int_0^u du'\,\mathcal{V}'(q+B(u'))\,B(u-u')\right]\,.$$
(VI.E.5)

This equation represents the homogeneized version of Eq.(VI.E.1). It is important to note that it *is not homogeneous in time* and thus its solutions are not either. The homogeneity holds at the level of non linear functions of the correlation functions and not at the level of times. It has no free parameters. $q$ and $x$ are fixed by Eqs. (III.23), (VI.A.1) and (VI.A.2). The invariance of the aging regime equations under time-reparametrisations $t \to h(t)$ implies the invariance of Eqs.(VI.E.4) and (VI.E.5) under dilatations of $u$: $u \to \eta u$.

Equation (VI.E.5) can be solved exactly for some special choices of $\mathcal{V}$ such as the $p$-spin spherical model and the RSGM in the limit of vanishing mass. The solutions are $B(u) = 2q_{PS}(1-\exp(-\eta u))$ and $B(u) = \eta u$, respectively. We shall describe these cases in detail at the end of this Section. In what follows we shall instead analyse the small and large $u$ behaviour of $B(u)$ for the general potential correlation.

We solve Eq.(VI.E.4) around $u \sim 0$, i.e. $b \sim q$, that corresponds to $B(u) \sim 0$, by first expanding in powers of $B(u)$ and then proposing a series expansion of $B(u)$ in terms of $u^\alpha$. $\alpha$ is a positive exponent that should be determined by the equation. An exponent $\alpha$ smaller than one is related to an irregularity at $u \sim 0$.

The expansion around $B(u) \sim 0$ yields

$$0 = \sum_{n=2}^{\infty}\left[\frac{\mathcal{V}^{(n+1)}(q)}{\mathcal{V}''(q)\,n!}(B(u))^n + \frac{2xq}{T}\frac{\mathcal{V}^{(n)}(q)}{(n-1)!}\frac{\partial}{\partial u}\int_o^u du'\,(B(u'))^{n-1}\,B(u-u')\right]\,. \quad (\text{VI.E.6})$$

It is easy to see that the zero-th and first order equations in $B(u)$ are immediately satisfied. The second order equation is however non-trivial:

$$0 = \frac{\mathcal{V}'''(q)}{\mathcal{V}''(q)} (B(u))^2 + \frac{4xq}{T} \mathcal{V}''(q) \frac{\partial}{\partial u} \int_0^u du' \, B(u') B(u - u') \, . \qquad \text{(VI.E.7)}$$

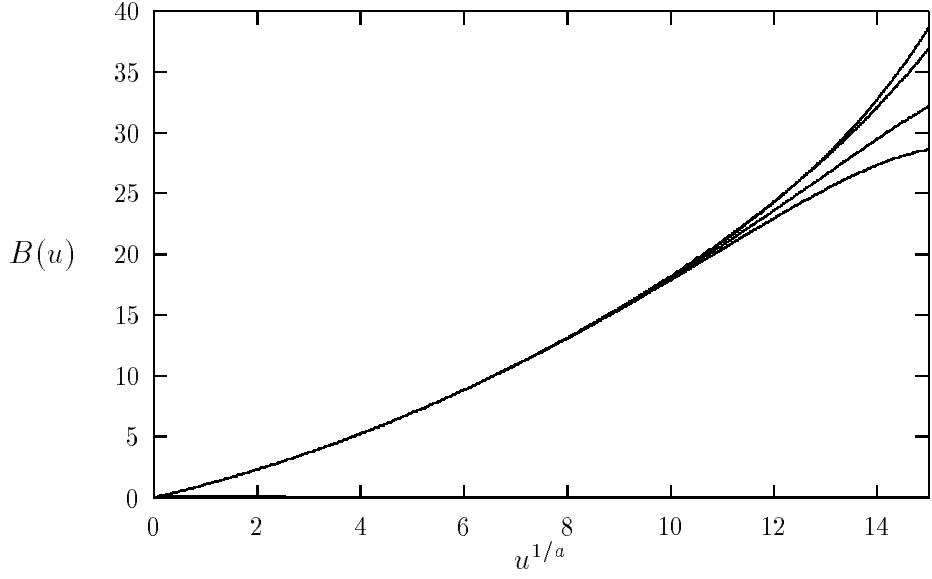

FIG. 4. $B(u)$ vs. $u^{1/a}$ with $1/a = \alpha$ $\gamma = 1.5$, $\theta = 2.5$, $T = 0.2$ and $\mu = 0$. From top to bottom the number of coefficients are $7, 9, 10, 8$.

We now proceed by proposing a formal series expansion for the function $B(u)$ around $u \sim 0$ in terms of $u^\alpha$

$$B(u) = \mathcal{B}(u^\alpha) = \sum_{m=0}^{\infty} b_m u^{m\alpha} \qquad \text{(VI.E.8)}$$

and then by solving Eq. (VI.E.6) term by term in the double series expansion. More precisely, Eq.(VI.E.7) at first order in $u^\alpha$ implies

$$0 = \left( \frac{\mathcal{V}'''(q)}{\mathcal{V}''(q)} + \frac{4xq}{T} \mathcal{V}''(q) \frac{(\Gamma(1+\alpha))^2}{\Gamma(1+2\alpha)} \right) (b_1 u^\alpha)^2 \, . \qquad \text{(VI.E.9)}$$

and this equation fixes the function

$$\rho(\alpha) \equiv \frac{(\Gamma(1+\alpha))^2}{\Gamma(1+2\alpha)} \qquad \text{(VI.E.10)}$$

and hence the exponent $\alpha$. The limiting values of $\rho$ are

$$\lim_{\alpha \to \infty} \rho(\alpha) = 0,$$
$$\rho(1) = 1/2$$
$$\lim_{\alpha \to 0} \rho(\alpha) = 1, .$$

For a general $\mathcal{V}$

$$\rho(\alpha) = -\frac{T}{4xq} \frac{\mathcal{V}'''(q)}{(\mathcal{V}''(q))^2} . \tag{VI.E.11}$$

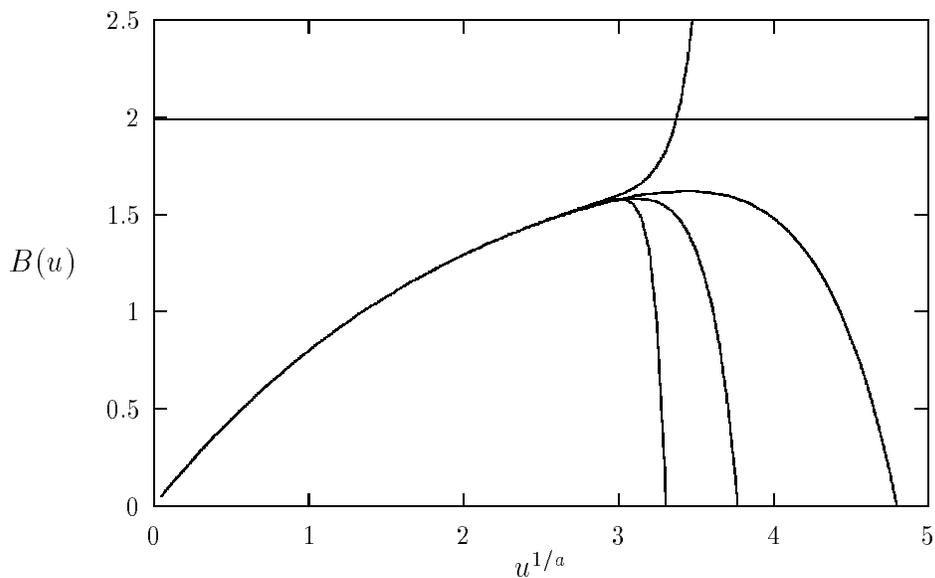

FIG. 5. $B(u)$ vs. $u^{1/a}$ where $1/a = \alpha$; $\gamma = 1.5$, $\theta = 2.5$, $T = 0.2$ and $\mu = 0.2$. The curve pointing upwards has 30 coefficients while the curves pointing downwards have $50, 20, 10$ from left to right. The straight line is $B = b_o - q$.

The first coefficient $b_1$ is not fixed as a consequence of the reparametrisations invariance. In Appendix D we describe how we can obtain the coefficientes $b_m, m \geq 2$, in terms of $b_1$ and then reconstruct the series for $B(u)$. In Figs. 4 and 5 we show $B(u)$ for model (1.1) with $\gamma = 1.5$, $\theta = 2.5$ and $T = 0.2$. Fig. 4 corresponds to the massless case and Fig. 5 to a massive case with $\mu = 0.2$. The different curves in both figures correspond to having approximated the series for $B(u)$ with a sum with different numbers of terms. We see that both in the massless and massive cases $B(u)$ can be obtained in this simple way up to reasonable values of $u^\alpha$. We have checked however that in the massive case a Padé approximation gives $B(u \to \infty) \sim b_o - q$ as expected. In the massless case we obtain the analytic asymptotic behaviour of $B(u)$ in the next section and find that it is in good accord with the tendency of the curves in Fig. 4.

### F. Solution for $b \sim b_o$, large $u$ behaviour for the massless models

The analysis of Eq.(VI.E.4) or (VI.E.5) for large $u$ is delicate and we shall not pursue it in all generality. We shall describe instead what happens in the limit of a vanishing mass for a general potential correlation.

In the limit $\mu \to 0$ Eqs.(VI.A.1), (VI.A.3) and (VI.A.2) imply

$$b_o \mu \sim -\frac{1}{x} \qquad \mathcal{V}'(q) - \mathcal{V}'(b_o) \sim -\frac{T}{2qx} \qquad \mu \tilde{q} \sim \frac{q\mathcal{V}'(q)}{T} - \frac{T\mathcal{V}(q)}{q\mathcal{V}'(q)} \qquad (\text{VI.F.12})$$

Thus $b_o \to \infty$, $\mathcal{V}'(b_o) \to 0$ and $\mathcal{V}'(q) \sim -T/(2qx)$. Using this result Eq.(VI.E.4) simplifies:

$$\mathcal{V}'(q) - \mathcal{V}'(q + B(u)) = -\frac{\mathcal{V}''(q)}{\mathcal{V}'(q)} \frac{\partial}{\partial u} \int_0^u du' \, \mathcal{V}'(q + B(u')) \, B(u - u') \, . \qquad (\text{VI.F.13})$$

Let us now study the asymptotic behaviour of $B(u)$ at large $u$. Since $B(u) \to \infty$ one must have

$$\int_0^u du' \, \mathcal{V}'(q + B(u')) \, B(u - u') \sim \frac{\mathcal{V}'(q)^2}{\mathcal{V}''(q)} u \qquad (\text{VI.F.14})$$

Defining $I \equiv \int_0^\infty du' \, \mathcal{V}'(q + B(u'))$ and assuming $I$ to be a convergent integral immediately implies $B(u) \sim \eta u$ with $\eta I = \frac{\mathcal{V}'(q)^2}{\mathcal{V}''(q)}$. One can check that this is the only possibility for short-range models. It implies

$$b(t, t') \sim q + \ln\left(\frac{h(t)}{h(t')}\right) \qquad (\text{VI.F.15})$$

in the limit $h(t')/h(t) << 1$ for *all* short-range models. This form which turns out to be exact for the RSGM in the massless limit (see next section) for all $u$, implies that $r(t, t')$ is only a function of $t'$ for widely separated times. We have checked this result numerically. In Fig. 6. we plot the response function $R(t, t')$ vs. the smaller time $t'$ for four times $t$ in the massless case. The parameters are $T = 0.2$, $\theta = 2.5$.

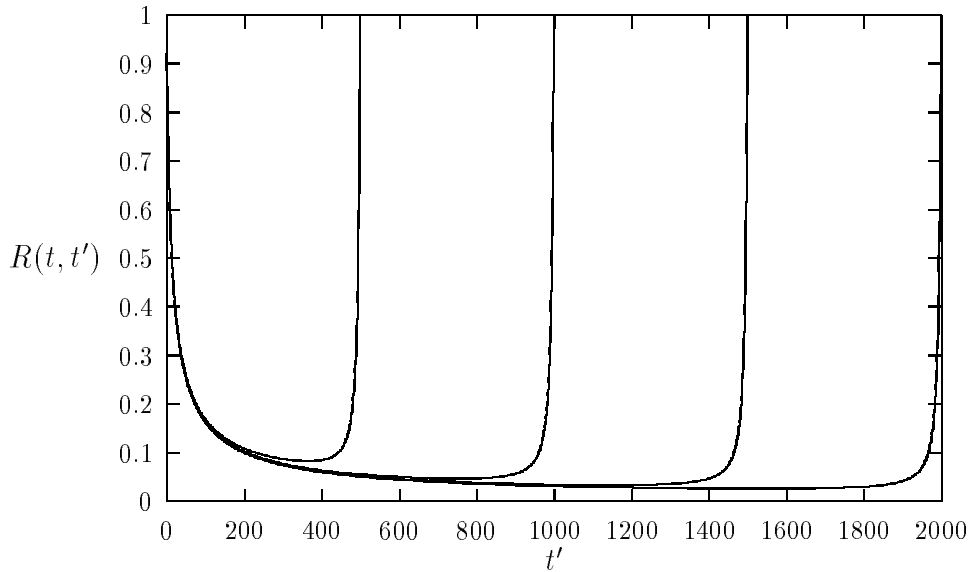

FIG. 6. The response function $R(t, t')$ vs. $t'$ for $t = ih$, $h = 0.2$ and $i = 100, 200, 300, 400$. $\mu = 0, T = 0.2, \gamma = 1.5$

In the present Section we are letting $\mu \to 0$ and using all the equations including (VI.A.1). Note that an identical equation (and solution) for $B(u)$ would be obtained by setting $\mu = 0$ first and then setting $b_o = \infty$ abandoning the equation (VI.A.1).

A very interesting quantity to calculate for the short-range models in particular in the massless limit is the asymptotic energy-density

$$\mathcal{E}_\infty(\mu \to 0) = \mu \tilde{q} + \frac{1}{T} \left( \mathcal{V}(0) - \mathcal{V}(q) \right) - 2x\mathcal{V}(q) . \qquad \text{(VI.F.16)}$$

As we shall see in the following Subsection, the asymptotic energy-density of the massive power law model with short-range correlations (1.5) is higher than the equilibrium energy-density obtained with the replica analysis[17]. Moreover, within the dynamic approach we find a non-trivial solution when $\mu \to 0$ with a finite asymptotic energy-density while statically the solution does not exist in the massless limit since $T_c(\mu \to 0) \to 0$ in the statics. This difference between asymptotic and equilibrium values seems to be characteristic of models solved by 1RSB ansatz statically, and it is related to the fact that the dynamic $B(X)$ is not equal to static Parisi function $b(u)$.

### G. Some particular models

The power law model

For model (1.1) $\rho(\alpha)$ becomes

$$\rho(\alpha) = -\frac{T}{xq}\frac{1+\gamma}{2\gamma}\,(\theta+q)^{\gamma}\,. \qquad (\text{VI.H.1})$$

In Fig. 2 we show the analytical prediction (broken straight lines) for this model with $\gamma = 1.5$, $\mu = 0.2$, $T = 0.2$ and $\theta = 2.5$. The full lines are the numerical solution to the dynamical equations for $t = 200, 300, 400$. from Eqs.(III.23)-(VI.A.2) $q = 1.17$ and $b_o = 3.17$. The FDT decay and $q$ are quite rapidly reached by the numerical solution. Instead, the convergence towards the aging theoretical curve and $b_o$ is slower. Note that the end of the full lines is at $b(t,0)$ that is slightly larger than $b_o = 3.17$, as expected.

When the mass $\mu$ tends to zero the asymptotic values $q$ and $b_o$, and $x$ can be easy obtained. They are given by the following equations

$$q^2\,(\theta+q)^{-(1+\gamma)} = \frac{2}{\gamma}\,T^2 \qquad b_o\mu \sim -\frac{1}{x} \sim \frac{q}{T}\,(\theta+q)^{-\gamma} \qquad (\text{VI.H.2})$$

$$\mu\,\tilde{q} \sim \frac{T(\theta+q)}{q}\left(\frac{1-2\gamma}{\gamma(1-\gamma)}\right)\,. \qquad (\text{VI.H.3})$$

When $\mu \to 0$, $b_o \to \infty$ and $\tilde{q} \to \infty$ in such a way that $b_o\mu$ and $\tilde{q}\mu$ remain finite. Thus, in the zero mass limit

$$\rho(\alpha) \sim \frac{1+\gamma}{2\gamma} \,. \qquad \text{(VI.H.4)}$$

Reconsidering the limiting behaviour of $\rho$, *cfr.* Eqs. (VI.E.11), we see that

$$\begin{aligned}
\gamma \to \infty &\quad\Rightarrow\quad \rho \to 1/2 \quad\Rightarrow\quad \alpha \to 1 \\
\gamma \to 1 &\quad\Rightarrow\quad \rho \to 1 \quad\Rightarrow\quad \alpha \to 0 \\
\gamma < 1 &\quad\Rightarrow\quad \rho > 1 \quad\Rightarrow\quad \alpha \not\exists
\end{aligned} \qquad \text{(VI.H.5)}$$

For $\gamma < 1$ (long-range potential) $\rho > 1$ but, from its definition (VI.E.10), $\rho(\alpha)$ cannot be greater than 1. Hence, we have demonstrated that the long-range potentials of type (1.5) do not have a one blob solution, at least in the massless limit, for any temperature. This demonstration can be extended to show that indeed the long-range model does not admit a one blob solution even when the mass is finite.

At the zero temperature limit we can obtain the explicit dependence of $q$, $b_o$ and $x$ on the parameters $\theta$ and $\gamma$:

$$\begin{array}{llll}
q \sim T\sqrt{\frac{2}{\gamma}} \theta^{\frac{1+\gamma}{2}} \ll 1 & & x \sim -\sqrt{\frac{\gamma}{2}} \theta^{\frac{\gamma-1}{2}} & = O(1) \,. \\
b_o \mu \sim \sqrt{\frac{2}{\gamma}} \theta^{\frac{1-\gamma}{2}} = O(1) & & \tilde{q}\mu \sim \frac{1}{\sqrt{2\gamma}} \theta^{(1-\gamma)/2} \left(\frac{1-2\gamma}{1-\gamma}\right) & = O(1)
\end{array} \qquad \text{(VI.H.6)}$$

Note that $x$ remains finite at zero temperature.

The asymptotic energy-density for the power law model is

$$\mathcal{E}_\infty = \mu \tilde{q} + \frac{1}{T} \frac{1}{2(1-\gamma)} \left(\theta^{1-\gamma} - (\theta+q)^{1-\gamma}\right) - x \frac{1}{(1-\gamma)} \left((\theta+q)^{1-\gamma} - (\theta+b_o)^{1-\gamma}\right) \,;$$
$$\text{(VI.H.7)}$$

in the massless case it reads

$$\mathcal{E}_\infty(\mu \to 0) = \frac{T}{\gamma q}(\theta + q) + \frac{1}{2T(1-\gamma)} \left(\theta^{1-\gamma} - (\theta+q)^{1-\gamma}\right) \,. \qquad \text{(VI.H.8)}$$

In Fig. 7 we plot the energy-density decay for the massless power law model with parameters $\gamma = 1.5$, $\theta = 2.5$ and $T = 0.2$. Equation (VI.H.8) implies $\mathcal{E}_\infty \sim -.137$. In the inset we plot $\ln(\mathcal{E}(t) + .137)$ *vs.* $\ln t$. One can see that the energy-density approaches its asymptotic value with a power law.

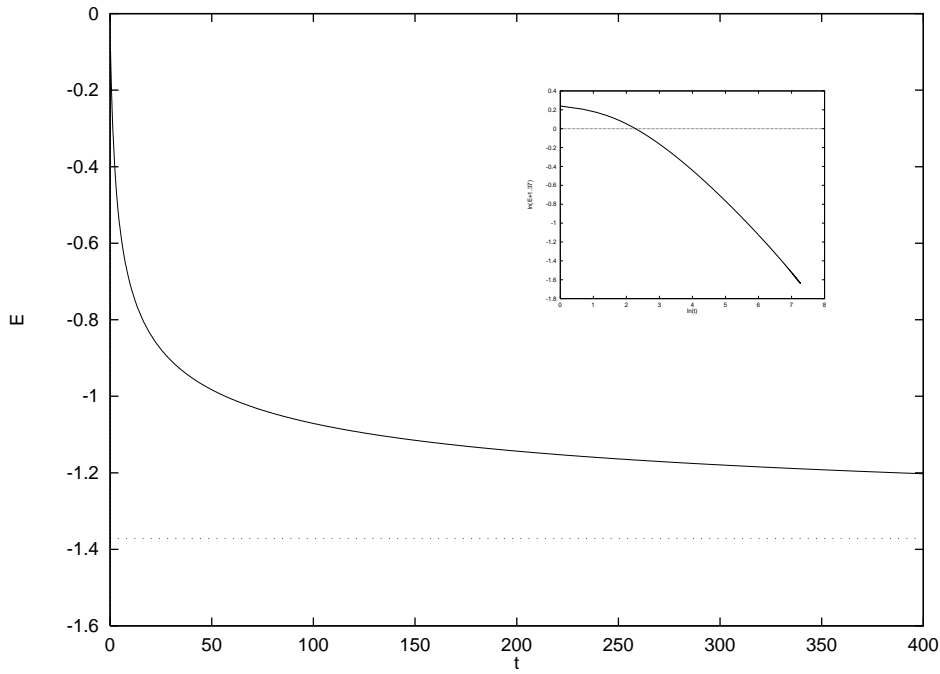

FIG. 7. Energy-density decay: $\mathcal{E}(t)$ vs. $t$ for the short-range model with $\gamma = 1.5$, $\theta = 2.5$, $T = 0.2$. In the inset $\ln(\mathcal{E}(t) + .137)$ vs. $\ln t$.

Finally, at zero temperature we have

$$\mathcal{E}_\infty(\mu \to 0, T \to 0) = \sqrt{\frac{2}{\gamma}} \theta^{(1-\gamma)/2} \ . \qquad \text{(VI.H.9)}$$

The zero mass random Sine Gordon model

In the limit of zero mass, Eqs. (VI.A.1) and (VI.A.2) can be easily solved. The maximum displacement $b_o$ tends to infinity in this limit but the product $\mu b_o$ stays finite; we obtain

$$q^2 \exp(-q/2) = \frac{4T^2}{\Delta} \qquad \mu b_o \sim -\frac{1}{x} \sim \frac{4T}{q} \qquad \tilde{q}\mu \sim \frac{4T}{q} \ , \qquad \text{(VI.H.10)}$$

Therefore, $\rho(\alpha)$ satisfies

$$\rho(\alpha) = \frac{1}{2} \quad \Rightarrow \quad \alpha = 1 \ . \qquad \text{(VI.H.11)}$$

Besides, this is a specially simple problem that can be solved completely from the original equations. Indeed, the solution is

$$b(t,t') = q + \ln\left(\frac{h(t)}{h(t')}\right) \qquad r(t,t') = \frac{q}{4T}\frac{h'(t')}{h(t')} \qquad \text{(VI.H.12)}$$

The exponent $\alpha = 1$ is consistent with this solution since it translates into

$$B(u) = \eta u \, . \qquad \text{(VI.H.13)}$$

The triangular relation function is thus $f(x,y) = x + y - q$. Note that a particular solution corresponds to $h(t) = t^\delta$ which we speculate is a good candidate for being the solution selected by matching at small times.

The $p$-spin spherical model

It is interesting to recall the known results for this model and compare them with those obtained with the general treatment described above. For spherically constrained models $b_o = 2$. In the case of the $p$-spin spherical model, with $p \geq 2$, Eq. (III.23) implies $p(p-1)/2\,(1-q_{PS})^2\, q_{PS}^{p-2} = T^2$. The mass $\mu$ in Eqs. (VI.A.1) and (VI.A.2) has to be interpreted as the constant long-time limit of the time-dependent $\mu(t)$ related to the Lagrange multiplier enforcing the spherical constraint, $cfr$. Eq. (2.7). Solving for $x$ one gets $x = -1/(2T)\,(p-2)(1-q_{PS})/q_{PS}$ These equations coincide with those found in Ref.[10] Eq.(VI.E.4) is solved, for all $u$, by $B(u) = 2q_{PS}(1 - \exp(-\eta u))$ with $\eta$ undetermined. Hence the function $\jmath$ defined in Eq.(IV.9) that relates any three displacements in the aging regime is $\jmath(b) = q_{PS}(1 - b/2)$ and the triangle relation is just $f(x,y) = xy/q_{PS}$.

If we study Eq.(VI.E.5) for $u \sim 0$ as described in the previous section, from Eq.(VI.E.11) $\rho(\alpha)$ is

$$\rho(\alpha) = \frac{1}{2} \quad \Rightarrow \quad \alpha = 1 \quad \text{for} \quad p > 2 \, . \qquad \text{(VI.H.14)}$$

The case $p = 2$ is particular. Both the third derivative in the numerator of Eq. (VI.E.11) and $x$ are zero. One cannot naively use Eq.(VI.E.11) to determine $\alpha$. Equation (VI.E.5) can be used also to check that the large $u$ analysis of it is delicate.

## H. From short-range models to long-range models: The appearance of the ultrametric solution

Let us study in detail the limit $\gamma \to 1$ from above for the power law model. In general, we have solved the short-range models with an ansatz consisting of a constant $x$ and a triangular relation $f(x,y) = \jmath^{-1}(\jmath(x)\jmath(y))$, with $\jmath^{-1}(x)$ a non-analytic function at $x=1$:

$$\jmath^{-1}(1+\epsilon) = q + c\epsilon^\alpha + O(\epsilon^{2\alpha}), \qquad (\text{VI.I.1})$$

that is different for different models, *i.e.* different $\mathcal{V}$. This implies for $f$

$$f(q+z, q+z') \propto (z^{1/\alpha} + z'^{1/\alpha})^\alpha + O(z^2, z'^2, zz'). \qquad (\text{VI.I.2})$$

In Section VI-H we have also shown when $\gamma$ tends to one the exponent $\alpha$ for the power law model tends to 0. Hence, when $\gamma \to 1$

$$f(q+z, q+z') \sim q + \max(z, z') + h.o.t \qquad (\text{VI.I.3})$$

and we see explicitly how the ultrametric solution is approached when the potential correlation range is increased.

The FDT theorem violation $x$ stays constant when $\gamma \to 1$ from above and this is also valid when the limit is approached from below, as can be seen from Eq.(VII.11).

# VII. SOLUTIONS TO THE DYNAMICAL EQUATIONS: LONG-RANGE POTENTIALS

We shall apply the ultrametric ansatz to Eqs.(V.3) and (V.4). The ultrametric ansatz means

$$b = f(b', b'') = \max(b', b'')$$
$$b'' = \overline{f}(b', b) = \max(b', b)$$

$\forall\, b, b', b''$ inside the interval $[q, b_o]$. Note that $\partial \overline{f}(b', b)/\partial b = 1$ and $\partial \overline{f}(b', b)/\partial b' = 0$, since $b > b'$. Furthermore $\partial^{(k)} \overline{f}(b', b)/\partial b^{(k)} = \partial^{(k)} \overline{f}(b', b)/\partial b'^{(k)} = 0$, $\forall k \geq 1$. In the dynamic equations (V.3) and (V.4) only $\overline{f}$ enters. We can then take variations of these equations w.r.t. $b$ safely.

Inserting the ultrametric form of $\overline{f}$ in Eq.(V.3) and in its variation with respect to $b$, and using Eq.(V.5) yields

$$0 = \frac{2q}{T}\left(\mathcal{V}''(q)F[b] + \frac{1}{4}\overline{M}[b]\right) - F[b]\overline{M}[b],$$
$$0 = X[b]\left[\frac{2q}{T}\left(\mathcal{V}''(q) - \mathcal{V}''(b)\right) - 4\overline{M}[b] + 4\mathcal{V}''(b)F[b]\right]. \qquad (\text{VII.1})$$

Solving these equations for $F[b]$ and $\overline{M}[b]$ we obtain

$$F[b] = \mp \frac{q}{2T}\left[\frac{\mathcal{V}''(b) - \sqrt{\mathcal{V}''(b)\mathcal{V}''(q)}}{\mathcal{V}''(b)}\right]$$
$$\overline{M}[b] = \pm \frac{2q}{T}\left[\mathcal{V}''(q) - \sqrt{\mathcal{V}''(b)\mathcal{V}''(q)}\right] \qquad (\text{VII.2})$$

The second variation of Eq. (V.3) combined with the above expresions for $F[b]$ and $\overline{M}[b]$ gives

$$X[b] = \mp \frac{q}{4T} \frac{\mathcal{V}'''(b)}{\mathcal{V}''(b)} \sqrt{\frac{\mathcal{V}''(q)}{\mathcal{V}''(b)}}. \qquad (\text{VII.3})$$

The sign must be chosen in such a way that $X[b]$ is negative; it depends on the relative signs of $\mathcal{V}'''$ and $\mathcal{V}''$. In the following we shall consider $\mathcal{V}'' < 0$ and $\mathcal{V}''' >$. This is the case for the power law model with long-range correlations. We shall then choose the lower signs in the expressions above. If $\overline{f}$ is a maximum, all the integrals appearing in the higher variations of the equation w.r.t. $b$ in the limit $b \to q$ vanish.

We require that $|X|$ be a decreasing or constant function of its argument $b$. This indeed selects the models that can be solved by an ultrametric ansatz. Imposing $|X[b]| < |X[q]|$ for $b > q$ yields

$$1 > \frac{\eta(b)}{\eta(q)} \quad \text{with} \quad \eta \equiv \frac{\mathcal{V}'''}{(-\mathcal{V}'')^{3/2}} \,. \tag{VII.4}$$

For instance, if the function $\mathcal{V}$ is of the form (1.5), we have

$$\left(\frac{\theta + q}{\theta + b}\right)^{\frac{1-\gamma}{2}} \leq 1 \,. \tag{VII.5}$$

If $\gamma > 1$ this inequality cannot be satisfied. This means that model (1.1) with short-range correlations cannot be solved by an ultrametric ansatz. Instead, if $\gamma \leq 1$, the condition is satisfied for all $b$ and model (1.1) with long-range correlations can be solved by an ultramtric ansatz.

As a second example we can examine the RSGM model. Eqs. (1.9) and (VII.4) imply

$$\frac{\exp(b/4)}{\exp(q/4)} \leq 1 \tag{VII.6}$$

Since $b \geq q$ then this imposes $q = b$ and the RSGM does not admit an ultrametric solution.

Finally, we study the $p$-spin model. In this case, $1 - B/2 = C$ and the correlation function decays from $q$ when $t' \to t$ to 0 when $t' \to 0$. Then,

$$\left(\frac{q}{c}\right)^{p/2} \leq 1 \tag{VII.7}$$

Since $q \geq c$ the inequality is not satisfied.

In conclusion, from the particular cases we presented in the Introduction, only model (1.1) with long-range interactions, *i.e.* $\gamma \leq 1$, admits an ultrametric solution. More generally, Eq.(VII.4) selects the potentials that admit an ultrametric solution.

The formula (VII.3) for $X$ together with the ultrametric form of $f$ represent the full solution for the model. Due to the time-reparametrisation invariance we have included in the equations we cannot go beyond these expressions and obtain the explicit time-dependence of the displacement $b$. We can check that Eq.(III.17) is also solved by this ansatz.

We can now explicitly obtain $q, b_o$ and $\tilde{q}$ and compare them with the static results obtained with the replica approach[5,17]. Equation (III.23) determines $q$ and with $X[b]$ given by Eq. (VII.3) after a bit of algebra we obtain $b_o$ from Eq. (V.5):

$$\mathcal{V}''(b_o) = -\frac{\mu^2}{4} \ . \tag{VII.8}$$

For the power law model with $\gamma \leq 1$ $q$ and $b_o$ are given by

$$\frac{q^2}{(\theta+q)^{1+\gamma}} = \frac{2T^2}{\gamma} \qquad (\theta+b_o)^{-(\gamma+1)/2} = \frac{\mu}{\sqrt{2\gamma}} \ . \tag{VII.9}$$

and $X[b]$ is

$$X[b] = -\frac{1}{2} \frac{(\gamma+1)}{\sqrt{2\gamma}} (\theta+b)^{(\gamma-1)/2} \ . \tag{VII.10}$$

In the limit $\gamma \to 1^-$ we have

$$X[b] = x = -\frac{1}{\sqrt{2}} \tag{VII.11}$$

that coincides with the constant $x$ obtained for the short-range models when $\gamma \to 1^+$

One can check that these equations for $q$ and $b_o$ coincide with the equilibrium equations obtained with a full replica symmetry breaking ansatz[17,12]. One can also check that the expression we obtained for $X(B)$ is formally identical to the Parisi function $x(q)$ of the replica treatment. Hence, the asymptotic dynamical values, in particular the energy-density, coincide with the static ones though, the particle *is not* visiting any equilibrium state.

In Fig. 1. we confront the analytical predictions (broken curves) to the numerical solution to the dynamical equations plotting $m(t,t')$ vs. $b(t,t')$ for $t = 200, 300, 400$. $\gamma = .5$, $T = 0.2$, $\mu = 0.2$ and $\theta = 2.5$ in the figure. $q$ and the $FDT$ relaxation converge rather quickly to the predicted value and straight line behaviour. However, the aging regime is still very far from the asymptotic analytical curve (for this choice of parameters the curve for the aging regime is unfortunately very similar to a straight line). The convergence for the long-range model is slower than for the short-range model (see Fig. 2 and the discussion in Section VI.G). We observe that the evolution of the numerical curves as $t$ increases is, though exceedingly slow, towards the dotted line.

This family of long-range models behaves in every respect as the Sherrington-Kirkpatrick model of a spin-glass[11].

## A. The phase diagram

The discussion of the phase diagram for power law models with $\gamma < 1$ is very different from the short range case. The function $q^2 \mathcal{V}''(q)$ is increasing monotonously thus there is no $T_c^{(1)}$. The critical temperature $T_c(\mu)$ is determined by: the condition $b_o = q$ that implies $\mathcal{V}''(2T_c^{(2)}/\mu) = -\mu^2/4$ (see Eq.(VI.B.4)). We can check that below this line $|X[b]| < 1/2T$ since:

$$-2TX(q) = -\frac{q}{2}\frac{V'''(q)}{V''(q)} = \frac{\gamma+1}{2}\frac{q}{\theta+q} < 1 \qquad \text{(VII.1)}$$

The dynamical phase diagram is thus identical to the statics.

# VIII. SUMMARY AND CONCLUSIONS

We have studied analytically the out of equilibrium dynamics of a particle moving in a general random potential. We have described in more detail three particular potential correlations associated to three well-known models: a power-law correlation, the Random Sine-Gordon model and the $p$-spin spherical spin-glass model.

The model with power-type long-range correlation (model (1.1)) is solved by an ultrametric ansatz for the triangular relations. In this case the asymptotic values of the one-time parameters such as the energy density, the equivalent of the Edwards-Anderson parameter $q$, the maximum displacement $b_o$, etc. coincide with the equilibrium values obtained with the full replica symmetry breaking scheme. Though the system never reaches equilibrium these quantities approach the equilibrium values asymptotically. This has been also shown to hold for the Sherrington-Kirkpatrick model and it is expected to hold in models without a threshold in the TAP energy landscape[10–12]. This seems to be a characteristic of models that are solved statically with a full replica symmetric scheme. Our results also agree with the numerical observations of Ref.[12].

Model (1.1) with short-range correlations is solved with a one blob ansatz for the triangular relations. This means that in the aging regime the function $X[b]$ measuring the departure from the FDT theorem is simply a constant and that the dynamical equation determines the function $\overline{f}$ relating any three correlation functions. We have found that in most of the models we consider the inverse function $\overline{f}$ presents an irregularity at the begining of the aging regime. This irregularity is a new and interesting feature that, although generic, has not been previously observed in the off-equilibrium dynamics of mean-field models (*e.g.*, it does not appear in the $p$-spin spherical model). A related irregularity, at the beginning of the out of equilibrium relaxations, has been predicted to occur in a model of traps with a broad distribution of trapping times by Bouchaud. This allows for a comparison of the mean-field glassy dynamics with the trap description of aging.

The one step trap model for the out of equilibrium relaxation of glassy systems[20] predicts that the integrated response of a system to a perturbation applied during the interval $[0, t_w]$ (the thermoremanent magnetisation for spin systems) has a very sharp decay to $q$ in a microscopic characteristic time. Afterwards the aging regime sets on and the waiting time $t_w$ is its only characteristic time. The decay is then described by $m(t, t_w) \sim (1-q)g(t) + qf(t/t_w)$ where the first term is the FDT relaxation and the second one is the aging relaxation. The aging part is characterised by two exponents

$$f\left(\frac{t}{t_w}\right) \sim 1 - \left(\frac{t}{t_w}\right)^{1-x_1} \quad \text{for } t \sim t_w \qquad f\left(\frac{t}{t_w}\right) \sim \left(\frac{t}{t_w}\right)^{x_2} \quad \text{for } t \gg t_w \quad \text{(VIII.1)}$$

Our results for short-range correlated potentials show that the mean-field dynamics of a particle in a random potential is also represented by a law of this type, at least in the beginning of the aging regime. Indeed when $(t-t')/t' = O(1)$ we find $m(t, t') = x(b(t, t') - b_o)$ which at the begining of this regime reads

$$m(t, t') \sim x\left(\frac{h(t)}{h(t')}\right)^\alpha + x(q - b_o) . \quad \text{(VIII.2)}$$

Thus, the exponent $x_1$ is related to our exponent $\alpha$.

At the end of the aging regime, *i.e.* when $t'/t \sim 0$, we can compare with the behaviour obtained in the limit of a massless particle:

$$b(t, t') \sim q + \ln\left(\frac{h(t)}{h(t')}\right) , \quad \text{(VIII.3)}$$

which naturally leads to logarithmic behaviour rather than power law at variance with (VIII.1).

The RSGM model with non-zero mass belongs to the family of models solved by a one blob ansatz with irregularities. The RSGM in the limit of a vanishing mass as well as the $p$-spin spherical spin-glass are particularly simple; they can be studied in an independent way and they serve as checks to the method to obtain the exponent characterising the decay at the beginning of the aging regime.

As regards the case of a general correlation $\mathcal{V}$ we have obtained the conditions that $\mathcal{V}$ must satisfy in order to be solved by each particular ansatz.

In many occasions it has been pointed out[40,30] that the mean-field spin-glass dynamical equations, with the assumption of the relaxation being at equilibrium, are very similar to Götze's mode coupling equations for a phenomenological description of the glass transition. The mode-coupling equations for glasses involve only the density-density correlation function and are time homogeneous by construction. Recently a very interesting first generalisation of the mode-coupling approach to account for non-equilibrium phenomena has been proposed[41]. The mean-field dynamical equations for the correlation and response function (without any time homogenous assumption) of the $p$-spin spherical spin-glass are indeed the dynamical equations arising from the mode-coupling treatment of a model without explicit quench disorder - the Amit-Roginsky $\phi^3$ model[41].

The equations we have studied in this paper constitute an enlargement of the class of possible extensions of the mode-coupling equations for glasses. Indeed, one can show that some of the mode-coupling equations of Götze are those associated to the high-temperature phases of the models we consider here with particular choices of the random potential. Though in these models there is explicit quench disorder it is however interesting that the same equations appear. The further advantage of these equations is that they imply a more complicated behaviour for their solutions as compared to the solution of the $p$-spin spherical spin-glass, for instance, they have singularities at the begining of the aging regime that may be relevant experimentally for glasses[42] This issue will be discussed in more detail in a separate publication[43].

Finally this work opens the way to study other disordered models in finite dimension. We expect some features found here to extend to higher dimensions, such as aging effects with irregularity in the scaling function. Recent results on instabilities of renormalization group flows in several models, such as the Sine-Gordon model[6], suggest that aging effects will be present.

# ACKNOWLEDGEMENTS

We want to especially thank J. Kurchan for several key suggestions as well as continuous encouragment, discussions and the careful reading of the manuscript. We also want to thank S. Franz and M. Mézard for lending us their numerical algorithm to solve the mean-field dynamical equations. During the course of this work we have also benefited from useful discussions with A. Bilal, J.P. Bouchaud, D. S. Dean, H. Horner, M. Mézard, J. Shapir. Finally, we want to thank the Vortex Phases Workshop Program at the Institute for Theoretical Physics, University of California at Santa Barbara, where this work was initiated. L. F. C. acknowledeges support from the Human capacity and mobility program of the European Community through the ccontract ERB4001GT933731.

# Appendix A

In this Appendix we extend the Gaussian Variational Method (GVM) to the dynamics of disordered systems. This method was previously used to study the statics[5]. Here we extend it by applying a Gaussian decoupling to the exact equations of motion obtained from the Martin-Siggia-Rose generating functional. We derive in this way a set of dynamical equations associated to the model defined by the Hamiltonian (1.1) with a general Gaussian random potential $V$.

We start by defining

$$-\delta(\boldsymbol{x} - \boldsymbol{x}') \Delta(\boldsymbol{\phi} - \boldsymbol{\phi}') \equiv \langle V(\boldsymbol{\phi}, \boldsymbol{x}) V(\boldsymbol{\phi}', \boldsymbol{x}') \rangle . \qquad (A.1)$$

The standard Martin-Siggia-Rose action entering the dynamical partition function $\mathcal{Z}_{\text{dyn}} = \int \mathcal{D}\boldsymbol{\phi}\, \mathcal{D}\hat{\boldsymbol{\phi}} \exp(-S)$ associated to a Langevin process is

$$S = \int d\boldsymbol{x} dt \sum_{\alpha}^{N} \left[ -T \, (i\hat{\phi}^{\alpha}(\boldsymbol{x}, t))^2 + i\hat{\phi}^{\alpha}(\boldsymbol{x}, t) \left( \partial_t - \nabla^2 + \mu \right) \phi^{\alpha}(\boldsymbol{x}, t) \right]$$
$$+ \int d\boldsymbol{x} dt dt' \sum_{\alpha\beta}^{N} \left[ -\frac{1}{2} i\hat{\phi}^{\alpha}(\boldsymbol{x}, t) \, \Delta''_{\alpha\beta}(\boldsymbol{\phi}(\boldsymbol{x}, t) - \boldsymbol{\phi}(\boldsymbol{x}, t')) \, i\hat{\phi}^{\beta}(\boldsymbol{x}, t') \right] , \qquad (A.2)$$

with $\Delta''_{\alpha\beta}(\boldsymbol{\phi} - \boldsymbol{\phi}') \equiv -(\delta^2/\delta\phi^{\alpha}\delta\phi'^{\beta})\Delta(\boldsymbol{\phi} - \boldsymbol{\phi}')$. The indices $\alpha, \beta$ label the components of the $N$-dimensional vectors $\boldsymbol{\phi}, i\hat{\boldsymbol{\phi}}$.

The dynamical equations for $R^{\alpha\beta}(\boldsymbol{x}, t; \boldsymbol{x}', t') \equiv \langle \phi^{\alpha}(\boldsymbol{x}, t) \, i\hat{\phi}^{\beta}(\boldsymbol{x}', t') \rangle$, $t \geq t'$, and $C^{\alpha\beta}(\boldsymbol{x}, t; \boldsymbol{x}', t') \equiv \langle \phi^{\alpha}(\boldsymbol{x}, t) \, \phi^{\beta}(\boldsymbol{x}', t') \rangle$ follow from

$$\left\langle i\hat{\phi}^{\beta}(\boldsymbol{x}', t') \frac{\delta S}{\delta i\hat{\phi}^{\alpha}(\boldsymbol{x}, t)} \right\rangle = \delta_{\alpha\beta} \, \delta^D(\boldsymbol{x} - \boldsymbol{x}') \, \delta(t - t') , \qquad (A.3)$$

$$\left\langle \phi^{\alpha}(\boldsymbol{x}', t') \frac{\delta S}{\delta i\hat{\phi}^{\beta}(\boldsymbol{x}, t)} \right\rangle = 0 , \qquad (A.4)$$

respectively. These are exact equations of motion which can be derived by standard methods[44]. The brackets here and in what follows denote a mean over the fields weighted with the dynamical effective action.

The first equation reads

$$\delta_{\alpha\beta}\,\delta^D(\boldsymbol{x}-\boldsymbol{x}')\,\delta(t-t') = \left(\partial_t - \nabla^2 + \mu\right)\,R^{\alpha\beta}(\boldsymbol{x},t;\boldsymbol{x}',t')$$
$$-\sum_\gamma \int dt''\,\left\langle i\hat{\phi}^\beta(\boldsymbol{x}',t')\,\Delta''_{\alpha\gamma}(\boldsymbol{\phi}(\boldsymbol{x},t) - \boldsymbol{\phi}(\boldsymbol{x},t''))\,i\hat{\phi}^\gamma(\boldsymbol{x},t'') \right\rangle \quad \text{(A.5)}$$

where we have used the fact that $\langle i\hat{\phi}i\hat{\phi}\rangle$ must be zero to preserve causality. The equation for $C$ reads

$$2T\,R^{\alpha\beta}(\boldsymbol{x}',t';\boldsymbol{x},t) = \left(\partial_t - \nabla^2 + \mu\right)\,C^{\alpha\beta}(\boldsymbol{x},t;\boldsymbol{x}',t')$$
$$-\sum_\gamma \int dt''\,\left\langle \phi^\alpha(\boldsymbol{x}',t')\,\Delta''_{\beta\gamma}(\boldsymbol{\phi}(\boldsymbol{x},t) - \boldsymbol{\phi}(\boldsymbol{x},t''))\,i\hat{\phi}^\gamma(\boldsymbol{x},t'') \right\rangle . \quad \text{(A.6)}$$

The last term in the r.h.s. of the above equations can be computed using a Gaussian approximation, *i.e.* assuming that the fields $\boldsymbol{\phi}$ and $i\hat{\boldsymbol{\phi}}$ have a Gaussian distribution. Using the Ito prescription that states $R(\boldsymbol{x},t;\boldsymbol{x},t)=0$, and the following rule for averaging any set of Gaussian variables $\{\varphi_i\}$ : $\langle \varphi_i F(\varphi)\rangle = \sum_j \langle \varphi_i\varphi_j\rangle \langle F'_j(\varphi)\rangle$, the equation for $R$ becomes

$$\delta_{\alpha\beta}\,\delta^D(\boldsymbol{x}-\boldsymbol{x}')\,\delta(t-t') = \left(\partial_t - \nabla^2 + \mu\right)\,R^{\alpha\beta}(\boldsymbol{x},t;\boldsymbol{x}',t')$$
$$-\sum_{\gamma\delta\epsilon}\int dt''\,R^{\delta\gamma}(\boldsymbol{x},t;\boldsymbol{x},t'')\left(R^{\epsilon\beta}(\boldsymbol{x},t;\boldsymbol{x}',t') - R^{\epsilon\beta}(\boldsymbol{x},t'';\boldsymbol{x}',t')\right)$$
$$\times \left\langle \Delta^{(4)}_{\alpha\gamma\delta\epsilon}(\boldsymbol{\phi}-\boldsymbol{\phi}')\right\rangle . \quad \text{(A.7)}$$

Similarly, the equation for $C$ reads:

$$2T\,R^{\alpha\beta}(\boldsymbol{x}',t';\boldsymbol{x},t) = \left(\partial_t - \nabla^2 + \mu\right)\,C^{\alpha\beta}(\boldsymbol{x},t;\boldsymbol{x}',t') - \sum_\gamma \int dt''\,R^{\alpha\gamma}(\boldsymbol{x}',t';\boldsymbol{x},t'')\langle\Delta''_{\beta\gamma}(\boldsymbol{\phi}-\boldsymbol{\phi}')\rangle$$
$$-\sum_{\gamma\delta\epsilon}\int dt''\,R^{\delta\gamma}(\boldsymbol{x},t;\boldsymbol{x},t'')\left(C^{\epsilon\alpha}(\boldsymbol{x},t;\boldsymbol{x}',t') - C^{\epsilon\alpha}(\boldsymbol{x},t'';\boldsymbol{x}',t')\right)$$
$$\times \left\langle \Delta^{(4)}_{\beta\gamma\delta\epsilon}(\boldsymbol{\phi}-\boldsymbol{\phi}')\right\rangle . \quad \text{(A.8)}$$

where $\langle ..\rangle$ denotes an average over a Gaussian distribution. These are the general dynamical equations obtained from the GVM. We now specialize to the case of O(N) symmetry, *i.e.* isotropy, which also includes $N=1$ (RSGM) as a trivial case, and show that with the definition of $\mathcal{V}$ appropriate for studying the large $N$ limit, the *same* function $\hat{\mathcal{V}}$ has to be used in the statics and the dynamics.

Defining the two functions:

$$-N \, \mathcal{V}\left(\frac{\phi^2}{N}\right) \equiv \Delta(\phi) \qquad -N \, \hat{\mathcal{V}}\left(\frac{\phi^2}{N}\right) \equiv \langle \Delta(\phi) \rangle \qquad (A.9)$$

Using isotropy the equations (A.7),(A.8) contain only $\Delta''_{\alpha\beta}$ and $\Delta''_{\gamma\gamma\alpha\beta}$ which are related to $\hat{\mathcal{V}}$ through:

$$\langle \Delta_{\alpha\beta}(\varphi) \rangle = \frac{1}{N}\delta_{\alpha\beta} \int \frac{d^N q}{(2\pi)^N} \, \Delta(q) q^2 \exp(-\frac{q^2 \varphi^2}{2N})$$

$$= 2\delta_{\alpha\beta}\hat{\mathcal{V}}'(\frac{\varphi^2}{N})$$

$$\langle \sum_\gamma \Delta_{\gamma\gamma\alpha\beta}(\varphi) \rangle = 4\delta_{\alpha\beta}\hat{\mathcal{V}}''(\frac{\varphi^2}{N}) \qquad (A.10)$$

Replacing these expressions in (A.7),(A.8) one recovers the dynamical equations (2.1),(2.2) with $\hat{\mathcal{V}}$ substituted for $\mathcal{V}$. In the limit of $N \to \infty$ these two functions become identical[5].

One can also apply the Gaussian decoupling in the exact equations of motion for the static replica theory. This provides an alternative way of obtaining the approximated saddle point equations of Ref.[5]. The exact saddle point equation is:

$$(\nabla^2 + \mu)\langle \phi_a(x)\phi_b(0) \rangle - \langle \phi_a(x) \sum_{c,d=1}^{n} \frac{\delta}{\delta \phi_b(0)} \left( \sum_{c,d=1}^{n} \mathcal{V}(\phi_c(0) - \phi_d(0)) \right) \rangle \qquad (A.11)$$

Applying now the Gaussian decoupling one recovers the saddle point equations of Ref.[5].

# Appendix B

In this Appendix we describe the derivation of the long-time dynamical equations, *i.e.* for times such that $t > t' \gg 1$ and $(t - t')/t' = 1$.

Neglecting the time-derivative, the $r$-Eq. reads:

$$\frac{\partial r(t,t')}{\partial t} \sim 0 \sim r(t,t') \left( -\mu + 4 \int_0^{t_-} ds\, \mathcal{V}''(B(t,s))\, R(t,s) + 4 \int_{t_-}^{t} ds\, \mathcal{V}''(b_F(t-s)) r_F(t-s) \right)$$
$$- 4 \int_{t'}^{t'_+} ds\, \mathcal{V}''(b(t,s))\, r(t,s)\, r_F(s-t') - 4 \int_{t'_+}^{t_-} ds\, \mathcal{V}''(b(t,s))\, r(t,s)\, r(s,t')$$
$$- 4 \int_{t_-}^{t} ds\, \mathcal{V}''(b_F(t-s))\, r_F(t-s)\, r(s,t') \,. \tag{B.1}$$

We have explicitly separated in the integrals the FDT regimes (integrals symbolically denoted $\int_{t_-}^{t} ds$) from the widely separated time regimes. In the last term of Eq. (B.1) $s$ varies from $t_-$ to $t$ and it is very far away from $t'$. The assumption is that in that case the functions vary very slowly (we have already neglected the time derivative in the l.h.s.). Hence $r(s, t')$ is almost constant in this interval and it can be approximated by $r(t, t')$. The last term then cancels the third one. In addition, using the same approximation and the FDT relations (III.5) one computes explicitly the integrals involving the FDT parts:

$$-4 \int_{t'}^{t'_+} ds\, \mathcal{V}''(b(t,s))\, r(t,s)\, r_F(s-t') \sim -4 \mathcal{V}''(b(t,t')) r(t,t') \int_0^{\infty} d\tau\, r_F(\tau)$$
$$\sim \frac{-2q}{T} \mathcal{V}''(b(t,t')) r(t,t')$$
$$4 \int_{t_-}^{t} ds\, \mathcal{V}''(b_f(t-s))\, r_F(t-s) = -\frac{2}{T} \left( \mathcal{V}'(0) - \mathcal{V}'(q) \right) \tag{B.2}$$

using $b_F(\infty) = q$ and $b_F(0) = 0$. Finally, the second term inside the parenthesis involves both short and long times since the integral goes from the initial time, that is strictly zero to $t_-$ that is large and belongs to the asymptotic regime. One should then separate the contribution from finite times to that of long times

$$\int_0^{t_-} ds\, \mathcal{V}''(B(t,s))\, R(t,s) \sim \int_0^{0_+} ds\, \mathcal{V}''(b_s(t,s))\, r_s(t,s) + \int_{0_+}^{t_-} ds\, \mathcal{V}''(b(t,s))\, r(t,s) \,. \tag{B.3}$$

We recall the weak long-term memory hypothesis and assume that the finite times do not contribute to the long-time dynamics and that the system forgets at large-times what happened in the very short times after the initial time.

Then the first term in the r.h.s. of Eq.(B.3) is neglected and one finally obtains the equation for the slow part of the response function:

$$0 = r(t,t') \left( -\mu + 4 \int_0^t ds \, \mathcal{V}''(b(t,s)) r(t,s) - \frac{2q}{T} \mathcal{V}''(b(t,t')) \right)$$
$$- 4 \int_{t'}^t ds \, \mathcal{V}''(b(t,s)) \, r(t,s) r(s,t') \tag{B.4}$$

The same procedure can be carried through for $b(t,t')$. First one can use that at large $t,t'$, $C(t,t) \simeq C(t',t') \simeq c_F(0) = \tilde{q}$. Then, the integrals which appear in the $b$-Eq. and contain FDT pieces are:

$$2 \int_{t_-}^t ds \, \mathcal{V}'(b_F(t-s)) \, r_F(t-s) = \frac{-1}{T} (\mathcal{V}(0) - \mathcal{V}(q))$$
$$-2 \int_{t'_-}^{t'} ds \, \mathcal{V}'(b(t,t')) \, r_F(t'-s) = \frac{-q}{T} \mathcal{V}'(b(t,t'))$$
$$2 \int_{t_-}^t ds \, \mathcal{V}''(b_F(t-s)) \, r_F(t-s) \, b_F(t-s) = \frac{q}{T} \mathcal{V}'(q) + \frac{1}{T} (\mathcal{V}(0) - \mathcal{V}(q))$$
$$2 \int_{t_-}^t ds \, \mathcal{V}''(b_F(t-s)) \, r_F(t-s) \, (b(t,t') - b(s,t')) \sim 0 \tag{B.5}$$

One also needs to compute:

$$-2 \int_{t'_-}^{t'+} ds \, \mathcal{V}''(b(t,s)) \, r(t,s) \, b_F(s,t') \sim -2 \mathcal{V}''(b(t,t')) \, r(t,t') \int_{t'_-}^{t'+} ds \, b_F(s,t') \, . \tag{B.6}$$

One can see that this integral is subdominant and gives a vanishing contribution to the $b(t,t')$ equation in the large $t$ limit. Since $r(t,t')$ itself is of order $1/t$ (see below) it results that the FDT part of this integral is of order $q(t'_+ - t'_-)/t$. The rapidly varying part of the total integral is thus vanishingly small compared to the total integral which is dominated by the slowly varying part.

Again, this equation could have contributions from the initial times $0 \leq s < 0_+$. We assume they are subleading, *i.e.*

$$0 \sim \int_0^{0_+} ds\, \mathcal{V}'(B(t,s))\, (R(t,s) - R(t',s))$$

$$0 \sim \int_0^{0_+} ds\, \mathcal{V}''(B(t,s))\, R(t,s)\, (B(t,s) + B(t,t') - B(s,t'))\,. \tag{B.7}$$

One finally obtains the equation for the slow part of the displacement correlation function:

$$\begin{aligned}
0 = &\left(-\frac{\mu}{2} + 2\int_0^t ds\, \mathcal{V}''(b(t,s))\, r(t,s)\right) b(t,t') + T + \frac{q}{T}(\mathcal{V}'(q) - \mathcal{V}'(b(t,t'))) \\
&+ 2\int_0^t ds\, \mathcal{V}'(b(t,s))\, r(t,s) - 2\int_0^{t'} ds\, \mathcal{V}'(b(t,s))\, r(t',s) \\
&+ 2\int_0^t ds\, \mathcal{V}''(b(t,s))\, r(t,s) b(t,s) - 2\int_0^{t'} ds\, \mathcal{V}''(b(t,s))\, r(t,s) b(t',s) \\
&- 2\int_{t'}^t ds\, \mathcal{V}''(b(t,s))\, r(t,s) b(s,t')\,.
\end{aligned} \tag{B.8}$$

From (2.3) one finds, using the same decomposition of the integrals and formulae (B.5) to evaluate the FDT parts, the large-time dynamical equation for $C(t,t)$ and $C(t,t')$.

# Appendix C

Let us also indicate the derivation of another (equivalent) form for the equation for $R(t,t')$ and $r(t,t')$. Starting from (2.1) one defines:

$$\mathcal{F}(t,t') = -\int_{t'}^{t} ds\, R(t,s) \tag{C.1}$$

*i.e.*, such that $R(t,t') = \frac{\partial F(t,t')}{\partial t'}$. One obtains:

$$\frac{\partial}{\partial t}\frac{\partial}{\partial t'}\mathcal{F}(t,t') = -\mu \frac{\partial \mathcal{F}(t,t')}{\partial t'} + 4\int_0^t ds\, \mathcal{V}''(B(t,s))\frac{\partial \mathcal{F}(t,s)}{\partial s}\frac{\partial \mathcal{F}(t,t')}{\partial t'}$$
$$-4\int_{t'}^{t} ds\, \mathcal{V}''(B(t,s))\frac{\partial \mathcal{F}(s,t')}{\partial t'}\frac{\partial \mathcal{F}(t,s)}{\partial s} \tag{C.2}$$

integrating over $t'$ and using $F(t',t') = 0$ one obtains:

$$\frac{\partial F(t,t')}{\partial t} = -1 + (-\mu + 4\int_0^t ds\, \mathcal{V}''(B(t,s))\frac{\partial F(t,s)}{\partial s})F(t,t')$$
$$-4\int_{t'}^{t} ds\, \mathcal{V}''(B(t,s))\, F(s,t')\frac{\partial F(t,s)}{\partial s} \tag{C.3}$$

The integration constant has been fixed using the limit $t \to t'$ and:

$$\lim_{t' \to t}\frac{\partial F(t,t')}{\partial t} = -\lim_{t' \to t}(R(t,t) + \int_{t'}^{t} ds\frac{\partial F(t,s)}{\partial s})$$
$$= -1 - \lim_{\epsilon \to 0}(r_F(0) - r_F(\epsilon)) = -1 \tag{C.4}$$

Note that at this stage equation (C.3) is exact and provides another convenient form to equation (2.1).

## Appendix D

In this Appendix we obtain the recursion relations for the function $B(u)$ suitable for a numerical solution. Defining $v = u^\alpha$ and $B(u) = vg(v)$, one introduces $g(v)^n = \sum_{m=0}^{\infty} g(n,m) v^m$ and $g(n,0) = 1$. Equation (VI.E.6) then leads to:

$$0 = \sum_{n=2}^{\infty} c_n v^n g(v)^n + X c_{n-1} v^n H(n,v) \tag{D.1}$$

where $c_n \equiv \mathcal{V}^{(n+1)}(q)/n!$, $X = -2Tx/q$ and

$$H(n,v) = \sum_{m=0}^{\infty} h(n,m) v^m = \frac{1}{v^n} \frac{d}{du} \left[ uv^n \int_0^1 dx\, x^{(n-1)/a} (1-x)^\alpha g(vx^\alpha)^{n-1} g(v(1-x)^\alpha) \right]. \tag{D.2}$$

After some algebra, we obtain two recursion relations for the coefficients $g(n,m)$ and $h(n,m)$:

$$g(n,m) = \sum_{j=0}^{m} g(n-1,j) g(1,m-j) \tag{D.3}$$

$$h(n,m) = \sum_{j=0}^{m} g(n-1,j) g(1,m-j) (1 + \frac{n+m}{a}) \beta(1 + \frac{n-1+j}{a}, 1 + \frac{m-j+1}{a}), \tag{D.4}$$

where $\beta(x,y) = \Gamma(x)\Gamma(y)/\Gamma(x+y)$. At the iteration $n$ the coefficients $g(i,j)$ and $h(i,j)$ with $i = 3, \ldots, n$, $j = n - i = 0, \ldots, n - 3$ are first directly obtained from the recursion relations (D.3) and (D.4). The coefficients $g(1, n-2)$, $g(2, n-2)$ (and $h(1, n-2)$, $h(2, n-2)$) are then obtained from the equation

$$2g(1,0) g(1, n-2) \left( c(2) + X c(1) \tilde{\beta}(1, n-1) \right) = - \sum_{k=3}^{n} \left( c(k) g(k, n-k) + X c(k-1) h(k, n-k) \right)$$

$$- \sum_{k=1}^{n-3} g(1,k) g(1, n-2-k) \left( c(2) + X c(1) \tilde{\beta}(1 + k, n - 1 - k) \right) \tag{D.5}$$

with

$$g(2, n-2) = 2g(1,0) g(1, n-2) + \sum_{1}^{n-3} g(1,k) g(1, n-2-k) \tag{D.6}$$

$$h(2, n-2) = 2g(1,0) g(1, n-2) \tilde{\beta}(1, n-1) + \sum_{1}^{n-3} g(1,k) g(1, n-2-k) \tilde{\beta}(1+k, n-1-k). \tag{D.7}$$

where $\tilde{\beta}(x,y) = (1 + (x+y)/a) \beta(1 + x/a, 1 + y/a)$.

# REFERENCES

[1] For a recent review see,. *e.g.* G. Blatter, M. V. Feigel'man, V. B. Geshkenbein, A. I Larkin and V.M. Vinokur; Rev. Mod. Phys. 66, 1125 (1994).

[2] J. Cardy and S. Ostlund; Phys. Rev. B25 6899 (1982).

J. Toner and D. di Vincenzo; Phys. Rev. B41 632 (1990).

Y. Y. Goldschmidt and B. Schaub; Nucl. Phys. B251, 77 (1985).

Y-C Tsai and Y. Shapir; Phys. Rev. Lett. 69, 1773 (1992). Phys. Rev. E50, 3546 (1994); *ibid*, 4445 (1994).

T. Hwa and D.S. Fisher; Phys. Rev. Lett. 72 2466 (1994).

[3] G.G. Batrouni and T. Hwa; Phys. Rev. Lett. 72, 4133 (1994).

D. Cule and Y. Shapir; Phys. Rev. Lett. 74, 114 (1995); *Numerical studies of the glass transition in the roughness of a crystalline surface with a disordered substrate*, prep. Rochester Univ. 1994.

E. Marinari, R. Monasson, J. Ruiz-Lorenzo; *How (super) rough is the glassy phase of a crystalline surface with a disordered substrate?*, cond-mat 9503074.

H. Rieger; *Comment on: dynamic and static properties of the randomly pinned flux array*, cond-mat 9503092.

[4] E. Medina, T. Hwa, M. Kardar and Y.-C. Zhang; Phys. Rev. A39, 3059 (1989).

[5] M. Mézard and G. Parisi; J. Phys. A23, L1229 (1990); J. Phys. I (France) 1, 809 (1991); *ibid* 2, 2231 (1992).

[6] P. Le Doussal and T. Giamarchi; Phys. Rev. Lett. 72, 1530 (1994); *ibid* 74, 606 (1995); *Elastic theory of flux lattices in presence of weak disorder*, cond-mat 9501087.

S. E. Korshunov, Phys. Rev. B48, 3969 (1993).

[7] L. C. E. Struik; '*Physical aging in amorphous polymers and other materials*', Elsevier, Houston (1978).

L. Lundgren, P. Svedlindh, P. Nordblad and O. Beckman; Phys. Rev. Lett. 51, 911 (1983).

E. Vincent, J. Hammann e M. Ocio; in *Recent progress in Random Magnets*, ed. D. H. Ryan, World Scientific, Singapore (1992),

[8] V. Dotsenko, M. Feigel'man and I. Ioffe; *Spin glasses and related problems* Soviet Scientific Reviews 15 (New York: Harwood Academic).

[9] H. Eissfeller and M. Opper; Phys. Rev. Lett. 68, 2094 (1992).

[10] L. F. Cugliandolo and J. Kurchan; Phys. Rev. Lett. 71, 173 (1993); Phil. Mag. B71 (1995).

[11] L. F. Cugliandolo and J. Kurchan; J. Phys. A27, 5749 (1994).
A. Baldassarri, L. F. Cugliandolo, J. Kurchan and G. Parisi; J. Phys. A28, 1831 (1995).

[12] S. Franz and M. Mézard; Europhys. Lett. 26, 209 (1994); Physica A209, 1 (1994).

[13] L. F. Cugliandolo and D. Dean; *Full dynamical solution of a mean-field spin-glass model*, cond-mat 9502075, to be published in J. Phys. A; *On the dynamics of a spherical spin-glass model in a magnetic field*, Saclay preprint, 1995.

[14] J. Villain, B. Semeria, F. Lancon and L. Billard; J. Phys. C16, 6153 (1983).

[15] U. Schulz, J. Villain, E. Brézin, H. Orland; J. Stat. Phys. 51, 1 (1988)

[16] L. F. Cugliandolo, J. Kurchan and P. Le Doussal; in preparation.

[17] A. Engel; Nucl. Phys. B410 [FS], 617 (1993).

[18] A. Crisanti and H-J Sommers; Z. Phys. B87, 341 (1992).

[19] J. M. Kosterlitz, D. J. Thouless and R. C. Jones; Phys. Rev. Lett. 36, 1217 (1976).

[20] J.-P. Bouchaud; J. Phys. I (France) 2, 1705 (1992).
J.-P. Bouchaud, E. Vincent and J. Hammann; J. Phys. I (France) 4, 139 (1994).
J.-P. Bouchaud and D.S. Dean; J. Phys. I (France) 5, 265 (1995).

[21] H. Sompolinsky and A. Zippelius; Phys. Rev. Lett. 45, 359 (1981); Phys. Rev. B25, 274

(1982).

[22] H. Sompolinsky; Phys. Rev. Lett. 47, 935 (1981).

[23] M. Mézard, G. Parisi and M. A. Virasoro; *Spin-glass theory and beyond*, World Scientific, Singapore (1987).

S. Franz and J. Kurchan; Europhys. Lett. 20, 197 (1992).

A. Baldassarri; Tesi di Laurea Univ. di Roma I, 1995 and in preparation.

[24] H. Kinzelbach and H. Horner; J. Phys. I (France) 3, 1329 (1993).

[25] H. Kinzelbach and H. Horner; J. Phys. I (France) 3, 1901 (1993).

[26] Y. G. Sinai; Theor. Probab. Its Appl. 27 247 (1982).

for review see J. P. Bouchaud, A. Comtet, A. Georges and P. Le Doussal; Ann Phys. 201 285 (1990) and Europhys. Lett. 3 653 (1987)

[27] R. Durrett; Comm. Math. Phys. 104, 87 (1986).

[28] M. Feigel'man and V. Vinokur; J. Phys. (France) 49, 1731 (1988).

[29] E. Marinari and G. Parisi; J. Phys. A26, L1149 (1993).

[30] A. Crisanti, H. Horner and H-J Sommers; Z. Phys. B92, 257 (1993).

[31] D. Cule and Y. Shapir; *Non-ergodic dynamics of the 2D random phase Sine-Gordon model: applications to vortex arrays and disordered substrate surfaces*, cond-mat 9410067.

[32] J. O. Andersson, J. Mattson and P. Svedlindh; Phys. Rev. B46, 8297 (1992); *ibid* B49, 1120 (1994).

H. Rieger; J. Phys. A26, L615 (1993); J. Phys. I (France), 883 (1994); *'Montecarlo studies of Ising spin-glasses and random field systems*, cond-mat 9411017 to appear in Annual Reviews of Computational Physics.

L. F. Cugliandolo, J. Kurchan and F. Ritort; Phys. Rev. B49, 6331 (1994).

[33] S. Ciuchi and F. De Pasquale; Nucl. Phys. B300 [FS], 31 (1988).

[34] M. Hazewinkel; *Formal groups and its applications* (New York: Academic).

[35] J. Kurchan; J. Phys. (France) 2, 1333 (1992).
L. F. Cugliandolo, S. Franz, J. Kurchan and M. Mézard; unpublished.

[36] T. Nieuwenhuizen; *Exactly solvable model of a quantum spin-glass*, cond-mat 9408056, to be published in Phys. Rev. Lett.

[37] D. J. Thouless, P. W. Anderson and R. Palmer; Phil. Mag. 35, 593 (1977).

[38] M. Potters and G. Parisi; *Mean-field equations for spin models with orthogonal interaction matrices*, cond-mat 9503009.
R. Monasson; *The structural glass transition and the entropy of the metastable states*, cond-mat 9503166

[39] W. Götze; in *Liquids, freezing and the glass transition*, J. P. Hansen, D. Levesque and J. Zinn-Justin eds. (Les Houches Session LI, Elsevier Sc. Publishers BV, 1991).

[40] H-J Sommers and K. H. Fischer, Z. Phys. B, 125 (1985).
T. R. Kirkpatrick and Thirumalai; Phys. Rev. B36, 5388 (1987).

[41] S. Franz and J. Hertz; Phys. Rev. Lett. 74, 2114 (1995).

[42] J-P Bouchaud; talk given in *Low temperature dynamics and phase-space structure of complex systems*, NORDITA, March 1995, unpublished.

[43] J-P Bouchaud, L. F. Cugliandolo, J. Kurchan and M. Mézard; in preparation.

[44] J. Zinn-Justin; *Quantum field theory and critical phenomena*, (Clarendon Press, Oxford) 1989.